\documentclass[letterpaper,12pt]{article}
\pdfoutput=1

\usepackage{jheppub}
\usepackage{amsmath}
\usepackage{graphicx}
\usepackage{subfig}
\usepackage{cancel}
\usepackage[dvipsnames]{xcolor}
\usepackage{color}
\usepackage{mathrsfs}

\def\({\left(}
\def\){\right)}
\def\[{\left[}
\def\]{\right]}

\newcommand{\reef}[1]{(\ref{#1})}

\newcommand{\pd}[2]{\frac{\partial#1}{\partial#2}}
\newcommand{\deriv}[2]{\frac{d#1}{d#2}}

\newcommand{\ban}[1]{\begin{align}#1\end{align}}

\newcommand{\gp}[1]{{g^{(2)}_{#1}}}
\newcommand{\mathbreak}{\notag\\}

\newcommand{\lc}{L_\text{crit}}

\arxivnumber{1605.02803}

\title{Adiabatic corrections to holographic entanglement in thermofield doubles and confining ground states}

\author{Donald Marolf}
\author{and Jason Wien}

\affiliation{Department of Physics, University of California, Santa Barbara, CA 93106, USA}

\emailAdd{marolf@physics.ucsb.edu}
\emailAdd{jswien@physics.ucsb.edu}

\abstract{We study entanglement in states of holographic CFTs defined by Euclidean path integrals over geometries with slowly varying metrics.  In particular, our CFT spacetimes have $S^1$ fibers whose size $b$ varies along one direction ($x$)  of an ${\mathbb R}^{d-1}$ base.  Such examples respect an ${\mathbb R}^{d-2}$ Euclidean symmetry.  Treating the $S^1$ direction as time leads to a thermofield double state on a spacetime with adiabatically varying redshift, while treating another direction as time leads to a confining ground state with slowly varying confinement scale.  In both contexts the entropy of slab-shaped regions defined by $|x - x_0| \le  L$ exhibits well-known phase transitions at length scales $L= L_{crit}$ characterizing the CFT entanglements.  For the thermofield double, the numerical coefficients governing the effect of variations in $b(x)$ on the transition are surprisingly small and exhibit an interesting change of sign: gradients reduce $L_{crit}$ for $d \le 3$ but increase $L_{crit}$ for $d\ge4$.  This means that, while for general $L > L_{crit}$ they significantly increase the mutual information of opposing slabs as one would expect, for $d\ge 4$ gradients cause a small decrease near the phase transition.  In contrast, for the confining ground states gradients always decrease $L_{crit}$, with the effect becoming more pronounced in higher dimensions.
 }

\begin{document}
\maketitle

%%%%%%%%%%%%%%%%%%%%%%%%%%%%%%%%%%%%%%%%%%%%%%%%%%%%%%%%%%%%%%%%%%%%%%%%%%%%%%%%
%%%%%%%%%%%%%%%%%%%%%%%%%%%%%%%%%%%%%%%%%%%%%%%%%%%%%%%%%%%%%%%%%%%%%%%%%%%%%%%%
%%%%%%%%%%%%%%%%%%%%%%%%%%%%%%%%%%%%%%%%%%%%%%%%%%%%%%%%%%%%%%%%%%%%%%%%%%%%%%%%
\section{Introduction}

Entanglement is a fundamental property of quantum systems.  Studying this entanglement can provide insights into the nature of quantum states, and in particular into the scale of their correlations. In the holographic context, entanglement of the dual CFT is of particular interest through its association with the Einstein-Rosen bridges of black holes \cite{maldTFD} and perhaps  more generally \cite{VanRaamsdonk:2010pw,Czech:2012be,Maldacena:2013xja} with the emergence of bulk spacetime.

Our goal here is to generalize the analysis of holographic entanglement away from the commonly-considered highly symmetric systems.  For $d=2$ CFTs, much can be done exactly using conformal transformations.  This fact lies behind the recent analysis \cite{Marolf:2015vma} of the CFT states dual to asymptotically-AdS${}_3$ mutli-boundary vacuum wormholes.  In particular, it was understood there that such states admit a simple description at high temperatures where the state can be well-approximated by a thermofield double (TFD) over most of the CFT spacetime, perhaps with adiabatic variations from one point to another. While a full analysis comparable to \cite{Marolf:2015vma} is difficult in higher dimensions, we show below that computations of entanglement in spatialy-varying holographic TFDs remains tractable in the adiabatic limit.

We also investigate how entanglement in ground states of ($d-1$)-dimensional confining theories is affected by slow variations of the confinement scale. The particular class of confining theories we consider are those given by compactifying a $d$-dimensional holographic CFT on an $S^1$ as in \cite{witten}.  Such CFT ground states are related to the above thermofield doubles, as both are given by cutting open Euclidean path integrals over geometries with $S^1 \times {\mathbb R}^{d-1}$ topology.  Roughly speaking, the thermofield double states are given by cutting open the $S^1$ factor, while ground states of confining theories are given by cutting open a direction of the ${\mathbb R}^{d-1}$.  The particular path integrals considered here will involve warped products of the $S^1$ over ${\mathbb R}^{d-1}$ in which the size $b$ of the $S^1$ varies slowly.  This gives in the first interpretation TFD states in spacetimes with spatially varying redshift, and in the second ground states of confining theories in which the confinement scale varies with position.

Since we are interested in holographic field theories, in all cases we will work directly with the dual gravitational description.  Our CFT path integrals are then interpreted as integrals over all ($d+1$)-dimensional asymptotically locally Anti-de Sitter (AlAdS) spacetimes with boundary geometries as above.  Section \ref{prelim} begins below by reviewing the Euclidean bulk geometries recently constructed in \cite{metricpaper} that are expected to describe the dominant AlAdS saddle points.  For simplicity, we allow $b$ to vary only along one Cartesian direction of the ${\mathbb R}^{d-1}$ space.  While such solutions can be constructed by Wick rotating the standard fluid-gravity correspondence \cite{FG,FGref1,FGref2} in the presence of a time-translation Killing field and an appropriate regularity condition at the bifurcation surface, it is more natural to follow \cite{metricpaper} and use the $U(1)$ symmetry to develop a related but different expansion based on standard Schwarzschild-like coordinates rather than the ingoing Eddington-Finkelstein black hole coordinates of \cite{FG,FGref1,FGref2}.

We then proceed to compute holographic entanglement.  Section \ref{deconfined} pursues the thermofield-double interpretation and computes the effect of varying $b$ on the Ryu-Takayanagi (RT) entropies of slabs  of thickness $2L$ that preserve ${\mathbb R}^{d-2}$ Euclidean symmetry on a surface fixed by a reflection of the $S^1$.  We include both the case of slabs contained in a single copy of the CFT and that of pairs of diametrically opposed slabs in each of the two CFTs.  We thus also compute the effect of varying $b$ on the mutual information in opposing slabs and on the critical value $L_{crit}$ of $L$ at which the mutual information becomes non-zero.
Section \ref{confined}  then studies the effect on RT entropies for analogous slabs with $S^1 \times {\mathbb R}^{d-3}$ symmetry on a surface fixed by reflecting one direction in the  ${\mathbb R}^{d-1}$.  Here the interesting feature is the effect on the value $L_{crit}$ at which the entangling surface changes topology from connected ($S^1 \times [0,1] \times {\mathbb R}^{d-3}$)  to disconnected (two copies of the $(d-1)$-disk).  Readers focused on the final results may wish to jump to sections \ref{ATFDphase} and \ref{GSCphase} where the phase transitions are discussed.   We close with some final discussion in section \ref{discuss}.  The special case $d=2$ is treated analytically in appendix \ref{2+1}, and we discuss some estimation of the numerical uncertainty in appendix \ref{error}.

\section{Preliminaries}
\label{prelim}

We wish to describe holographic entanglement in CFT states defined by path integrals over geometries with topology $S^1 \times {\mathbb R}^{d-1}$ and metrics of the form
\begin{equation}
\label{bndymetric}
ds^2_{CFT} = dx^2 + \delta_{ij} dy^i dy^j + \alpha_d^2 b^2(x) d\theta^2 \ \ ,
\end{equation}
where $\alpha_d = \frac{2^{1-2/d}}{d}$ and $i = 1, \dots, d-2$.  We  take $\theta$ to have $b$-independent period  $2\pi$.  The relevant states are constructed by slicing open the path integral along a co-dimension one surface that we identify as $\tau=0$ for some Euclidean time coordinate $\tau$.  To have a good translation to Lorentz signature, we require a $\mathbb Z_2$ reflection symmetry $\tau \rightarrow - \tau$.  One natural choice is to take $\tau=\theta$, in which case we in fact slice the path integral along the pair of surfaces $\theta =0, \theta = \pi$.  The result is an entangled state on a pair of CFTs which gives an adiabatic generalization of the well-known thermofield double state.  The exact time-translation symmetry means that the  state is in thermal equilibrium when viewed from the perspective of either CFT alone.  However, after Wick rotation to Lorentz signature the $x$-dependent metric factor $g_{\theta \theta}$ means that the state lives in a spacetime with $x$-dependent gravitational redshift. This equilibrium thus requires any local notion of temperature (such as that defined by the inverse Euclidean period) to be $x$-dependent as well.  This interpretation is equally valid in the special case $d=2$ in which there are no $y$ directions.

For $d\ge 3$, there is a second interpretation given by choosing $\tau$ to be some $y$ direction (say, $y^1$), so that our CFT lives on a spacetime with a compact spatial $S^1$. States of this theory are constructed by slicing the path integral along $y^1=0$. For small $b$ one may Kaluza-Klein reduce on this $S^1$.  And as discussed in \cite{witten}, one expects the result to exhibit confinement with a scale set by $b$.  So when $b$ varies, one may think of the result as a confining theory with a position-dependent confinement scale.

But with either interpretation, so long as $b$ varies slowly reasoning analogous to that of \cite{witten} implies the bulk path gravitational integral with boundary conditions given by \eqref{bndymetric} to be dominated by a Euclidean solution to Einstein's equation in which the $S^1$ factor pinches off in the bulk; i.e., there will be a Killing field $\partial_\theta$ that generates a $U(1)$ isometry with a fixed-point set of topology ${\mathbb R}^{d-1}$.

When the function $b(x)$ varies slowly, the construction of such solutions may be organized in a derivative expansion.  Here we write $b= b(\epsilon x)$ for some small parameter $\epsilon$. The details of this expansion were recently described in \cite{metricpaper}, where it was argued that for slowly-varying $b(x)$ the solution should be well-approximated by the zero-order ansatz
\ban{
ds^2=\frac{\ell^2} {z^2} \left[dz^2+  {\left(1+\frac {z^d}{b^d}\right)^{4/d}} \left( dx^2 + \delta_{ij} dy^i dy^j \right) + \alpha_d^2 b^2 {\left(1-\frac{z^d}{{b}^d}\right)^2 \left(1+\frac{z^d}{{b}^d}\right)^{\frac{4}{d}-2}} d\theta^2 \right], \label{AdS-Soliton}
}
where we take $\theta$ to have period $2\pi$ for all profiles $b(x)$. For the case $b=constant$, the ansatz \eqref{AdS-Soliton} gives the metric on the Euclidean planar AdS-Schwarzschild black hole (or, equivalently, on the Euclidean AdS soliton).  The full metric is then taken to be of the form
\ban{
ds^2 &= \frac{\ell^2}{z^2} \left( g^{(0)}_{AB} \, dx^A dx^B +\epsilon\, g^{(1)}_{AB}\,  dx^A dx^B +\epsilon^2 \, g^{(2)}_{AB} \, dx^A dx^B +\cdots \right),
}
where the corrections $g^{(n)}_{AB}$ are determined by solving Einstein's equation with appropriate boundary conditions at each order in an adiabatic expansion and $x^A=(z,x,y^i,\theta)$ ranges over all bulk coordinates and similarly for $x^B$. As shown in \cite{metricpaper}, the $O(\epsilon)$ correction $g^{(1)}_{AB}$ vanishes and, writing $g_{y^iy^j} = g_{yy} \delta_{ij}$,  the $O(\epsilon^2)$ correction is of the form
\begin{eqnarray}
\label{notation}
g^{(2)}_{xx}(z,x) &=&\left({b'(x)}\right)^2  {\sf g}_{xx}^{(b')^2}(z/b) +\left(b(x) {b''(x)}\right)
{\sf g}_{xx}^{(bb'')}(z/b)  , \cr
g^{(2)}_{yy}(z,x) &=& \left({b'(x)}\right)^2  {\sf g}_{yy}^{(b')^2}(z/b) +\left(b(x) {b''(x)}\right)
{\sf g}_{yy}^{(bb'')}(z/b) , \cr
g^{(2)}_{\theta\theta}(z,x) &=& \alpha_d^2 \left[ \left(b(x)\,b'(x)\right)^2 {\sf g}_{\theta\theta}^{(bb')^2}(z/b) +  b(x)^3\,b''(x) \
{\sf g}_{\theta\theta}^{(b^3b'')}(z/b) \right].
\end{eqnarray}
Here the notation makes explicit all dependence on $b(x)$; there can be no further implicit dependence hidden in form of the coefficient functions ${\sf g}_{xx}^{(b')^2}(z/b)$, etc.  These coefficient functions were evaluated numerically in \cite{metricpaper} with boundary conditions that ensure that the boundary metric remains \reef{bndymetric} and that the spacetime remains regular at the fixed point set of $\partial_\theta$ (with the period of $\theta$ taken to be $2\pi$ independent of $b(x)$).

Below, we use the results of \cite{metricpaper} to calculate $O(\epsilon^2)$ corrections to the holographic entanglement entropy.  We also make use of two further results from \cite{metricpaper}.  The first is  that, for $d>2$, in the adiabatic expansion the Fefferman-Graham representation of our metrics takes the form
\ban{
ds^2_{z \ll b} =\frac {\ell^2} {z^2} \left[ dz^2+   \alpha_d^2 \left(b^2 + \epsilon^2 \frac {b \,  b''} {d-1} z^2 \right)d\theta^2+ \left(1+ \epsilon^2 \frac{ b''}{b (d-1)}z^2\right)dx^2 \right.\mathbreak
\left. + \left(1-\epsilon^2 \frac{ b''}{b (d-2) (d-1)}z^2\right)dy^i dy_i + O(z^d,\epsilon^4) \right]. \label{FGapprox}
}
The special case $d=2$ is treated in appendix \ref{2+1}.  The second is that near the fixed point set of $\partial_\theta$ the metric takes the form
\ban{
ds^2=g_{RR}|_{R=0} \left(dR^2 +  R^2 d\theta^2\right)+ g_{XX}|_{R=0} dX^2 + g_{YY}|_{R=0} \sum_{i=1}^{d-2} dY^i dY^i + O(R^2),\label{regular}
 }
 with $g_{RR}|_{R=0}, g_{XX}|_{R=0}, g_{YY}|_{R=0}$ functions of $X$ alone,
in terms of coordinates $X,R$ that satisfy
\ban{
z&=(1-R) b-\epsilon^2\, \frac 13 16^{-1/d}b \left( {b'}^2+\frac 2{\alpha_d^2 {d}^2}\,  \left. \partial_z^2g_{\theta\theta}^{(2)}\right|_{z=b}\right)+O(\epsilon^3)\notag \\
x&=X+\epsilon \, 16^{-1/d}\, b \, b '  \left( R +\frac 1{2}  R^2-\frac 16(d-2) R^3   \right)+O(\epsilon^3, R^4). \label{regtrans}
}
The key point of \eqref{regular} is that it ensures the desired regularity at $R=0$ (where $\partial_\theta=0$).  In terms of the Fefferman - Graham coordinates this set is described by $z= \tilde b$ where
\ban{
\tilde b &= b - \frac{\epsilon^2}{2} \, b^2 \left. \partial_z \gp{\theta\theta}\right|_{z=b} \label{floor} \, .
}
This is the black hole horizon for the adiabatic thermofield double interpretation and the IR floor for the confining one.

\section{Adiabatic Thermofield Doubles}
\label{deconfined}

We begin with the adiabatic thermal field double (ATFD) states defined by slicing our CFT path integral along the surfaces $\theta =0, \theta = \pi$ fixed by the reflection symmetry $\theta \rightarrow -\theta$. It is convenient to denote the union of these two surfaces by $C_{CFT}$.   A slight generalization of the Ryu-Takayangi proposal \cite{rt1,rt2} then states that the von Neumann entropy of the CFT in some region ${\cal R}_{CFT} \subset C_{CFT}$ can be computed as follows.  First, find the dominant saddle for the corresponding bulk path integral.  One expects it to be invariant under a corresponding reflection, and that this reflection leaves fixed a co-dimension one surface that we may call $C_{bulk}$.  Now find the minimal-area surface $\Sigma$ within $C_{bulk}$  that i) intersects the asymptotically AdS boundary on a set corresponding to the boundary $\partial {\cal R}_{CFT}$  of ${\cal R}_{CFT}$ and ii) is homologous to ${\cal R}_{CFT}$ within $C_{bulk}$ \cite{head,fur06a}. Since the Lewkowycz-Maldacena argument \cite{Lewkowycz:2013nqa} for the Ryu-Takayanagi proposal applies equally well to this generalization,  we shall use it freely below\footnote{See \cite{Haehl:2014zoa} for a discussion of the homology constraint in the context of the Lewkowycz-Maldacena argument.}.  We also note that the above prescription is equivalent to using the the covariant Hubeny-Rangamani-Takayanagi conjecture \cite{Hubeny:2007xt} in the Wick-rotated Lorentz-signature solution\footnote{We thank Veronika Hubeny for pointing out that this follows from the maximin construction of \cite{Wall:2012uf}.  Since the RT surface is minimal on the Cauchy surface $C_{bulk}$, its area can be no larger than that of the maximin surface.  But the time-reversal symmetry means that the RT surface is also an extremal surface in the full spacetime.  It can therefore have area no smaller than the maximin surface, as the latter agrees with the area of the smallest extremal surface.}.

For simplicity, we consider slab-shaped regions ${\cal R}_{CFT}$ defined by conditions of the form $|x-x_0| \le L$, perhaps also restricted to one of the two boundaries ($\theta =0$ or $\theta=\pi$).  The symmetries then reduce the problem of finding the minimal surface to studying curves in the $z,x$ plane, with the area being proportional to the volume of the $y$ directions. For purposes of displaying a finite result we take the $y$ coordinates to range over a torus of finite volume $V$.    Since we are interested in the decompactified limit, we will always assume each cycle of the $y$-torus to have length much larger than both $b$ and $L$.  In particular, we assume  that the dominant bulk saddle will continue to be given by \eqref{AdS-Soliton}.

A technical issue is that the area nevertheless remains infinite due to the divergence of the metric \eqref{AdS-Soliton} at $z=0$. As usual, we must renormalize this quantity in order to present finite results.  Thus we define
\ban{
A_\text{ren} =\lim_{z_0\to0}\left( A_\text{bare}(z_0)+\sum_{\partial \Sigma}\, A_\text{ct}(z_0)\right), \label{renorm}
}
where $A_\text{bare}(z_0)$ is the area of the part of the surface with $z> z_0$ and where there is one counter-term contribution $A_\text{ct}(z_0)$ for each boundary of the minimal surface $\Sigma$.

The general theory of such divergences is explained in \cite{Graham:1999pm}, which shows that when the bulk is described by pure Einstein-Hilbert gravity (with no additional matter fields) one may use counter-terms determined by the boundary metric alone\footnote{Interestingly, this is not true in general; see \cite{StateDep}.}, {though these generally involve both the induced geometry on $\partial \Sigma$, the extrinsic curvature of $\partial \Sigma$ \cite{Jacobson:1993vj, Dong:2013qoa}, and even derivatives of such extrinsic curvatures \cite{Miao:2014nxa} in high enough dimensions}.  See also \cite{Taylor:2016aoi} for a recent discussion of such counter-terms and their relation to \cite{Lewkowycz:2013nqa}.

To find a useful explicit form for our $A_\text{ct}(z_0)$ , we first write the area functional as
\begin{equation}
A _\text{bare}= V\, \ell^{d-1} \int_\lambda \mathcal A_\text{bare} d\lambda
\end{equation}
with
\begin{equation}
\mathcal A_\text{bare} = g_{yy}^{\frac 12(d-2)} \left(\frac{z'(\lambda)^2}{z(\lambda)^2} + x'(\lambda)^2 g_{xx}\right)^{1/2} \label{Afunc}
\end{equation}
for any parameter $\lambda$ along the associated curve in the $z,x$ plane.

 Near $z=0$ it is useful to set $\lambda =z$ and assume an adiabatic expansion of the form
\begin{equation}
x(\lambda)= x^{(0)}(\lambda) + \epsilon \,x^{(1)}(\lambda) + \cdots.
\end{equation}
The behavior of $x^{(0)}$ near $z=0$ is determined by the minimal surface equation of motion at order $\epsilon^0$.  This may be written
\begin{eqnarray}
0&=&\left((d+1) z^d-(d-1) b^d\right) {x^{(0)}}{}'(z)-(d-1)  \left(b^d-z^d\right) \left(1+ z^d/b^{d}\right)^{4/d} \left({x^{(0)}}{}'(z)\right)^3 \cr &+& z\,  \left(b^d+z^d\right) {x^{(0)}}{}''(z).
\label{Ozeromin}
\end{eqnarray}
Equation \eqref{Ozeromin} admits a power series solution of the form
\ban{
{x^{(0)}}(z)&= c_0 +c_d z^{d}+ c_{2d} z^{2d}+ \cdots \label{x0expansion}
}
Indeed, the result takes the form \eqref{x0expansion} in any metric having the same non-zero coefficients in its Fefferman-Graham expansion.  Since $g^{(1)}_{AB} =0$, at order $\epsilon$ the ansatz \eqref{AdS-Soliton} continues to give the full metric.  Noting that the endpoint conditions $x(z=0) = x_0 \pm L$ are independent of $\epsilon$ then also gives
\ban{
x^{(1)}(z)= \tilde c_d z^{d} + O(z^{d+1}).
}
So near $z=0$ the area density \reef{Afunc} becomes
\ban{
\mathcal A_\text{bare}&= \frac 1{z^{d-1 }}+ \frac 12\frac {\epsilon^2} {z^{d-1}} (d-2)\gp{yy}+O(z^0,\epsilon^3),
}
as any factors ${x^{(0)}}'(z)$ or $x^{(1)}(z)$ are of order $z^d$ and give corrections that vanish as $z\rightarrow 0$.

Combining the Fefferman-Graham expansion of the second order metric correction \reef{FGapprox} with the results above we find
\ban{
\mathcal A_\text{bare} = \frac 1{z^{d-1}}-\frac{\epsilon^2}{2 (d-1) } \frac{b''}b \frac 1{z^{d-3}}+ O(z^0, \epsilon^3), \label{cts}
}
so we may choose
\ban{
\begin{array}{l cc}
A_\text{ct}=  V \, \ell^{d-1}\left[-\frac {1} {(d-2)} \frac 1 {z_0^{d-2}}+ \epsilon^2 \frac{1}{2\, (d-1) (d-4)} \frac{ b''} b \frac 1{z_0^{d-4}} \right] &\hspace{0.5cm}& d\neq 2,4
\end{array}
}
There are no explicit $O(\epsilon)$ counter-terms since $g^{(1)}_{AB}=0$. { One may check that this choice of counterterms precisely implements the covariant counterterm prescription of \cite{Taylor:2016aoi} to $O(\epsilon^2)$. Following this prescription, the counterterms in $d=4$ will include a logarithmic as well as a constant piece, and in $d=2$ we only have the logarithmic piece. These terms are given by }
\ban{
\begin{array}{l cc}
A_\text{ct}=  V\, \ell^{3} \left[ -\frac {1} {2} \frac 1 {z_0^{2}}-  \epsilon^2\frac{ 1}{6} \frac{b''}b \log (z_0/\ell)+ \epsilon^2 \frac 1{12} \frac{b''}b \right] \ \ ,&\hspace{0.5cm} & d=4 \\
\\
A_\text{ct}= \ell\, \log (z_0/\ell) \ \ , && d=2 \label{ct2}
\end{array}
}
where no factor of $V$ appears in $d=2$ because there are no $y$-directions.

For $d=3$, the second counter-term in \eqref{ct2} vanishes; we nevertheless find that including it in the manner explained below improves the convergence of our numerics.

In practice, we find it convenient to renormalize in the following way. Let $\mathcal A_\text{ct} = -\left. \partial_{z_0} A_\text{ct}\right|_{z_0=z}$. Then we can write
\ban{
A_\text{ct} = \int_{z_0}^{z_\text{max}} \mathcal A_\text{ct} \, dz + \left. A_\text{ct}\right|_{z_0=z_\text{max}} \,
}
for any $z_{max}$.  In particular, we can take $z_{max}$ to be the maximal value of $z$ on our bulk extremal surface.  The renormalized area \reef{renorm} can then be written
\ban{
A_\text{ren}&= \lim _{z_0\to 0} \int_{z_0}^{z_\text{max}}\left( V\, \ell^{d-1} \mathcal A_\text{bare} + \sum_{\partial \Sigma} \mathcal A_\text{ct}\right)dz + \sum_{\partial \Sigma} \left.A_\text{ct} \right|_{z_0=z_\text{max}}\, , \mathbreak
&= \int_{0}^{z_\text{max}}\left(  V\, \ell^{d-1}\mathcal A_\text{bare} + \sum_{\partial \Sigma} \mathcal A_\text{ct}\right)dz +\sum_{\partial \Sigma} \left.A_\text{ct} \right|_{z_0=z_\text{max}}\, . \label{nrenormA}
}
The integral in the second line now converges, and is more stable to compute numerically. The price we pay is having to add the constant term involving $z_\text{max}$.  For $d=3$, we find that including the second (vanishing!) counter-term in \eqref{ct2} in this way improves our numerical convergence.  This appears to be due to the fact that we perform these integrals by changing variables to integrate over $x$ instead of $z$, and that the above renormalization removes an (integrable) singularity in the integrand that arises from the associated factor of $z'(x)$.

We are now ready to compute the entropies of our slabs $|x-x_0| \le L$.  For slabs contained in a single boundary,  we know on general grounds that the minimal surface will remain close to the conformal boundary when $L \ll b$ while for $L \gg b$ it will track the horizon closely over almost all of the interval $|x-x_0| \le L$.  The transition between these behaviors is smooth.     But if we take our slab to contain the regions $|x-x_0| \le L$ on both the $\theta=0$ and $\theta = \pi$ boundaries one finds a well-known phase transition \cite{Morrison:2012iz,Headrick:2010zt,Hubeny:2007xt,Hartman:2013qma}  when passing from the regime $L \ll b_0$ to the regime $L \gg b$.  In the former case, the minimal surface consists of two copies of that found in the single-boundary case.   In the latter case the minimal surface again has two connected components, but each component then stretches from $\theta =0$ to $\theta = \pi$ while remaining localized near $x =x_0 \pm L$.  This is the only context in which the minimal surface reaches or passes through the fixed point set of $\partial_\theta$. In each case we find the general solution numerically below and compare it with analytic approximations for $L \ll b$ and $b(\epsilon\, b')^{-1} \gg L \gg b$.  We also provide results for the case $L \gg  b (\epsilon\, b')^{-1} \gg b$.  The effect on the phase transition itself is analyzed in section \ref{ATFDphase}.

\subsection{Entropy on a single boundary}
\label{ATFDsingle}

We begin with connected slab-shaped regions ${\cal R}_{CFT}$ of width $2L$ lying in a single boundary. For generic values of the parameters, numerical calculations are required to find the extremal surface.  But certain limiting behaviors can be studied analytically.  We treat these cases first, and then compare the results with numerical studies of the general case. In the rest of this section, we set $x_0=0$ without loss of generality.

\subsubsection{Analytically tractable limits}
\label{ABTZanalytics}

Our first special case will be the large $L$ limit, as  the fact that the minimal surface closely tracks the horizon in this regime makes it particularly easy to study. To leading order in $L$, the renormalized area is just the horizon area in the region $|x| \le L$.  Using the induced metric on the horizon found in \cite{metricpaper} gives
\begin{eqnarray}
S&= \frac{ V\, \ell^{d-1}}{4G}\int_{-L}^{L}{d x}\left[ \frac{2^{2-2/d}}{ b^{d-1}}+\epsilon ^2\, \frac{2^{1-{6}/{d}}}{ b^{d-1}}  \left(\left.(d-2)\, \gp{yy}\right|_{z=b}+\left. \gp{xx}\right|_{z=b}+{b'}^2\right) + O(\epsilon^4)\right] + \dots,\hspace{1cm}
\label{Lge}
\end{eqnarray}
where the $\dots$ represent terms that do not grow with $L$ when $b$ remains bounded.

For $L$ larger than or comparable to $b/(\epsilon \, b')$, nothing more can be said without choosing an explicit function $b(\epsilon x)$.  But in the regime $b/(\epsilon\, b') \gg L \gg b$ we may
define
$b_0= b(0)$, $b'_0 = \left.\partial_x b\right|_{x=0}$, and $b''_0 = \left.\partial_x^2 b\right|_{x=0}$ and use the expansion
\ban{\label{expb}
b = b_0 + \epsilon\,x\, b'_0  +\frac 12 \epsilon^2 \, x^2 \, b''_0 + O(\epsilon^3)
}
to simplify \eqref{Lge}. Writing $A_{\text{ren}} = A^{(0)}_{\text{ren}}   + \epsilon^2 A^{(2)}_{\text{ren}}  + \dots$, we find
\ban{\label{largeL0}
\left. A_{\text{ren}}^{(0)}\right|_{L \gg b_0} \sim 2^{3-2/d} \frac{V\, \ell^{d-1}}{b_0^{d-1}}L ,  \ \
\left.  A^{(2)}_\text{ren}\right|_{L \gg b_0} &\sim \frac{1}{3} 2^{2-\frac{2}d} (d -1) \frac{V\, \ell^{d-1}}{b_0^{d+1}} \left(d\, {b_0'}^2  -b_0 b_0''\right) L ^3,
}
where $\sim$ indicates that we have found only the leading behavior for $L \gg b_0$.  Here we were able to obtain an analytic expression at order $\epsilon^2$ because the $L^3$ term comes only from the $O(\epsilon^2)$ term in \eqref{expb} and thus can involve the metric only at order $\epsilon^0$ as given by \eqref{AdS-Soliton}.

For $L \ll b_0$ the minimal surfaces will be confined to $z \ll b_0$, so we can estimate their area by truncating the Fefferman - Graham expansion \reef{FGapprox} for the metric to some order in $z$. The Fefferman - Graham expansion for $d=2$ has a non-trivial contribution from the boundary stress tensor at order $z^2$, so we treat this case separately in appendix \ref{2+1}.

Consulting the expansion \reef{FGapprox}, we see that to zeroth order in the adiabatic expansion we have Poincar\'e AdS${}_{d+1}$. So for $d> 2$ we find
 \ban{
A^{(0)}_\text{ren}  &= -\frac{2 \pi ^{\frac{d-1}{2}}  }{d-2} \left(\frac{\Gamma \left(\frac{d}{2 (d-1)}\right)}{ \Gamma \left(\frac{1}{2 (d-1)}\right)}\right)^{d-1} \frac {V\, \ell^{d-1}} {L^{d-2}} + O(L^2). \label{A0FG}
}
This leading term reproduces the standard result for slabs in Poincar\'e AdS${}_{d+1}$ as derived in \cite{rt1}.

Since $d\theta=0$ on the surface of time reflection symmetry, the truncated induced metric \eqref{FGapprox} depends on $b$ only at order $\epsilon^2$ and there can be no $O(\epsilon)$ correction to the minimal surface or its area.  And the fact that the zero-order surface is minimal means that there is no correction at order $O(\epsilon^2)$ from the second-order displacement of the surface within the zeroth-order spacetime.  Thus the only $O(\epsilon^2)$ contribution comes from evaluating the change in the area along the zeroth-order minimal surface that comes from including the $O(\epsilon^2)$ parts of \eqref{FGapprox}. This correction can be computed from the integral representation of the hypergeometric function found in equation (15.6.3) of \cite{dlmf} and yields
\ban{
A^{(2)}_\text{ren} =& \frac{\pi ^{\frac{d}{2}-2} \, _2F_1\left(\frac{1}{2},-\frac{d-4}{2 (d-1)};\frac{d+2}{2 (d-1)};1\right) }{3  (d-4)} \left(\frac {\Gamma \left(\frac{d}{2 d-2}\right)}{\Gamma \left(\frac{1}{2 d-2}\right)}\right)^{d-4}\frac{b_0''}{b_0} \, \frac {V \, \ell^{d-1}} {L^{d-4}}  + O(L^{4}) \, ,\label{smallL}
}
for $d>2, d\neq 4$ and \ban{
\label{smallL4}
A^{(2)}_\text{ren}=& \left[\frac 1 {18} \left({ 5}+\log \left[\frac{\pi ^3 \, \Gamma \left(\frac{2}{3}\right)^6}{4 \, \Gamma \left(\frac{1}{6}\right)^6}\right]\right)- \frac 13 \log L \right] \frac {b_0''}{b_0} V \ell^3 + O(L^{4}) \hspace{2cm} {\rm for} \ d=4 \, .
}

\subsubsection{Numerics and comparisons}

We now consider general values of $L \ll 1/(\epsilon\, b').$  This allows us to again use \eqref{expb} so that the results can depend only on the parameters $b_0, b'_0$, and $b''_0$.  For $d\neq 2,4$ we write
 \begin{equation}
 A_\text{ren} =  \frac{V\,\ell^{d-1}}{ b_0^{d-2}} {\sf A}(L/b_0), \label{dimensionless}
 \end{equation}
 where the function form of ${\sf A}(L/b_0)$ is determined only by dimensionless combinations of $b$ and its derivatives.
 For $d=2$ and $d=4$ it is useful to subtract the logarithmic dependence on $\ell$ coming from the regularization scheme \eqref{ct2} and write
\ban{
\label{d24}
 A_\text{ren} &=  {\ell}\, {\sf A}(L/b_0) +\ell\, \log (b_0/\ell)\, , \hspace{3.3 cm} d=2\, \mathbreak
 A_\text{ren} &=  \frac{V\,\ell^{3}}{ b_0^{2}} {\sf A}(L/b_0) -\epsilon^2 \, V\,\ell^{3}\frac1 6 \frac {b_0''}{b_0} \, \log (b_0/\ell)\, . \hspace{1cm} d=4\,.
}
We may then use the adiabatic expansion to write
 \ban{
 {\sf A}(L/b_0) &= {\sf A}^{(0)}(L/b_0)+\epsilon\,  {\sf A}^{(1)}(L/b_0)+\epsilon^2\,  {\sf A}^{(2)}(L/b_0)+ O(\epsilon^3) \, .
 }
 Now, the correction $ {\sf A}^{(1)}(L/b_0)$ would have to be proportional to the first-order adiabatic parameter $b'_0/b_0$.  But the sign of this parameter changes under $x \rightarrow -x$ whereas the area must be invariant.  So there can be no correction at this order.  We thus consider only the second order corrections, which must be linear in the two dimensionless second-order adiabatic parameters $(b'_0)^2$ and $b_0 \,b''_0$; i.e., we have
 \begin{equation}
 {\sf A}^{(2)}(L/b_0) = (b'_0)^2 {\sf A}^{(b')^2}(L/b_0) +  b_0\,b''_0 {\sf A}^{(bb'')}(L/b_0),
 \label{genO2}
 \end{equation}
 with ${\sf A}^{(b'{}^2)}, {\sf A}^{(bb'')}$ having no further dependence on $b(x)$.
\begin{figure}[h!]
\centering
\subfloat[][$L=0.1\, b_0$]{\includegraphics[width=0.45\textwidth]{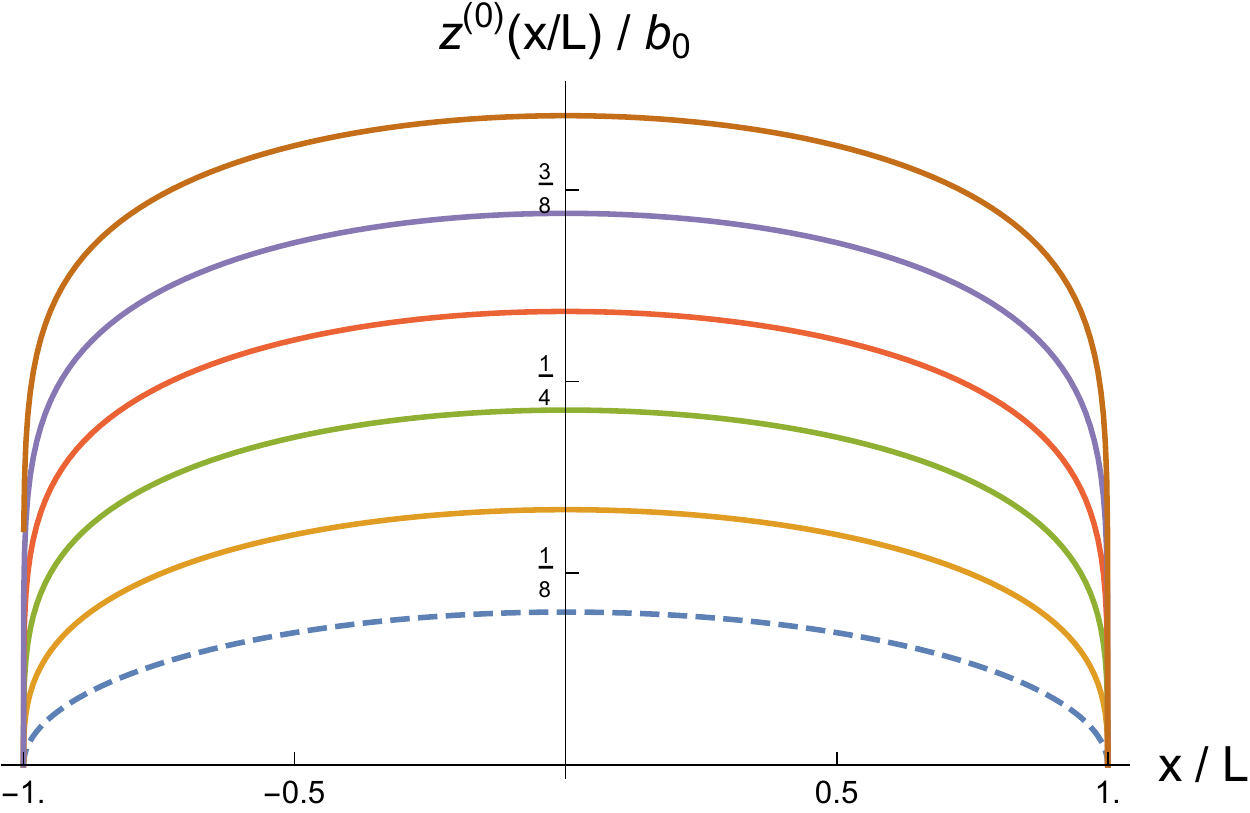}}\qquad
\subfloat[][$L=b_0$]{\includegraphics[width=0.45\textwidth]{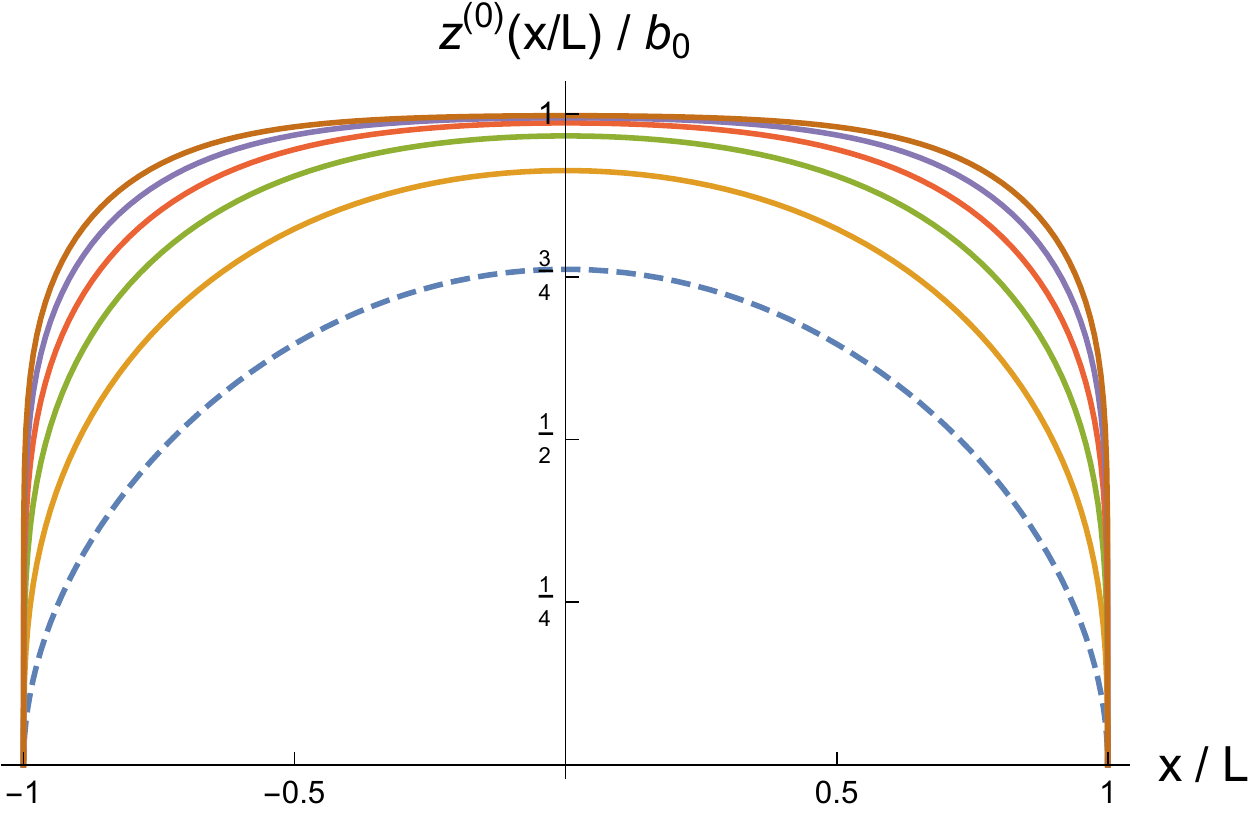}}\\
\subfloat[][$L=5 \, b_0$]{\includegraphics[width=0.55\textwidth]{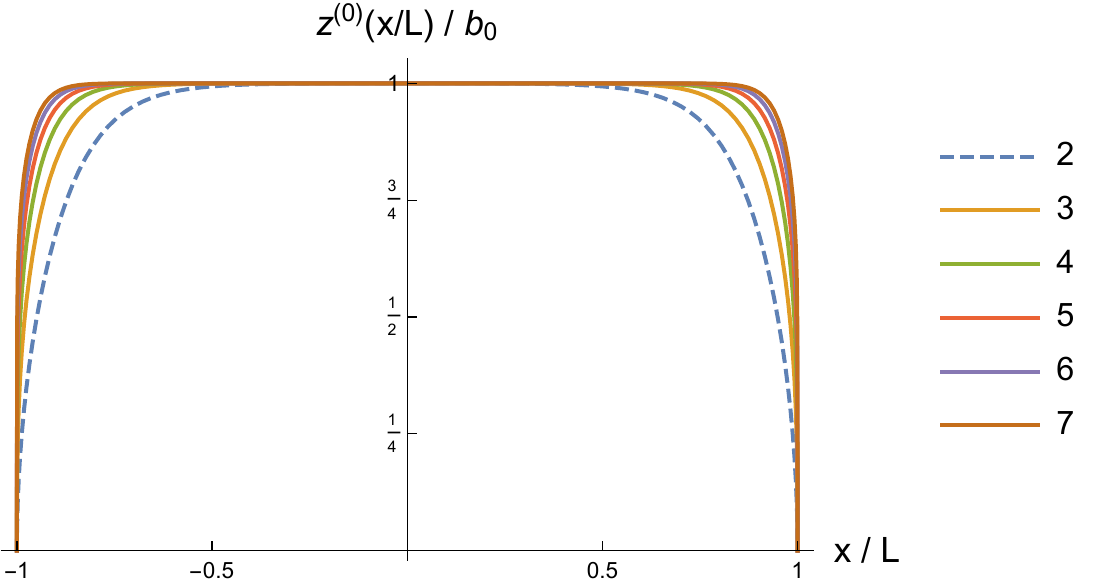}}
\caption{ Numerical solutions for $z^{(0)}(x/L)/b_0$ for slabs of width $2L$ on a single boundary with $2 \le d \le 7$. As $L$ increases from (a) to (c), the entangling surface quickly approaches the horizon as expected.
}
\label{surf0}
\end{figure}

\begin{figure}[h!]
\centering
\includegraphics[width=0.65\textwidth]{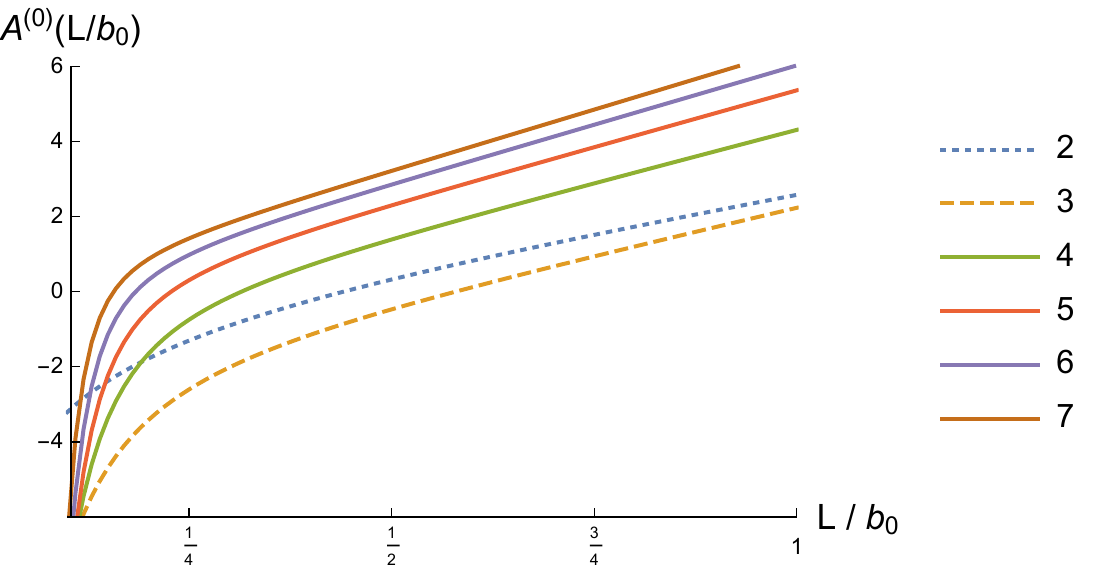}
\caption{ The rescaled zeroth order area ${\sf A}^{(0)}(L/b_0)$ for slabs of width $2L$ on a single boundary with $2 \le d \le 7$.  The curves interpolate between a power law proportional to $-(b_0/L)^{d-2}$ for $L \ll b_0$ and linear growth for $L \gg b_0$ where the entangling surface tracks the horizon closely. For $d=2$ the small $L/b_0$ behavior is logarithmic.
}
\label{A0}
\end{figure}

Even at order $\epsilon^0$ we require numerics to solve for the surface that extremizes the area \eqref{Afunc}.
We use the Newton-Raphson method outlined in \cite{numerics}. Figure \ref{surf0} shows the solution for $z^{(0)}(x/L)/b_0$ with $2 \le d \le 7$ and various interval sizes.  Results for the zeroth order area are shown in figure \ref{A0}.

\begin{figure}[h!]
\centering
\subfloat[][$L=0.1 \,b_0$]{\includegraphics[width=0.45\textwidth]{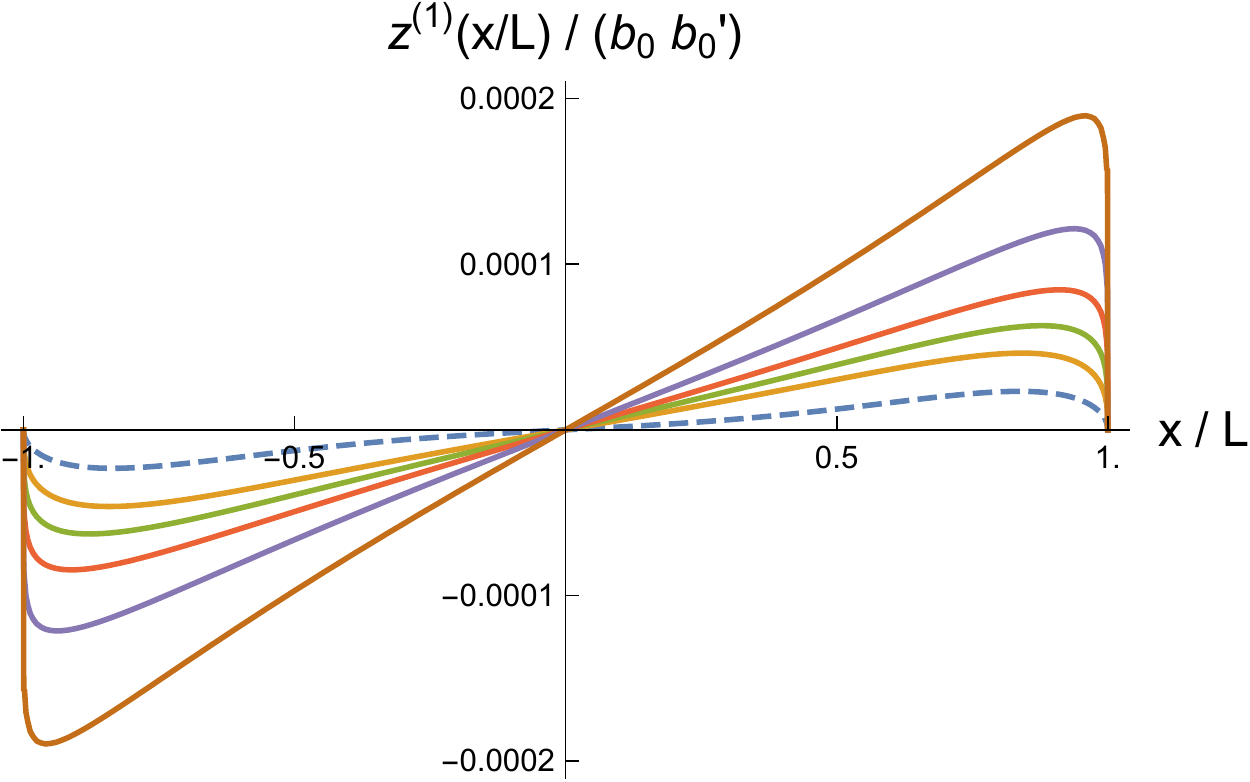}}\qquad
\subfloat[][$L=b_0$]{\includegraphics[width=0.45\textwidth]{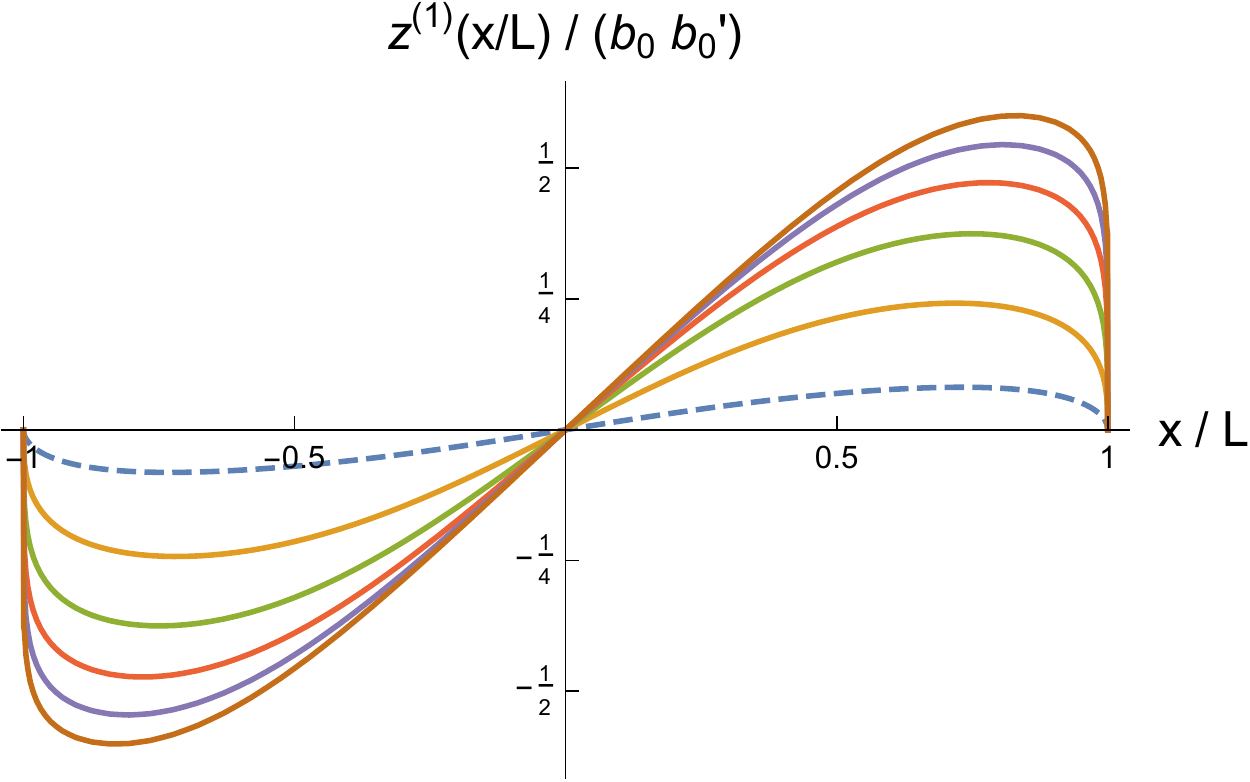}}\\
\subfloat[][$L=5\,b_0$]{\includegraphics[width=0.55\textwidth]{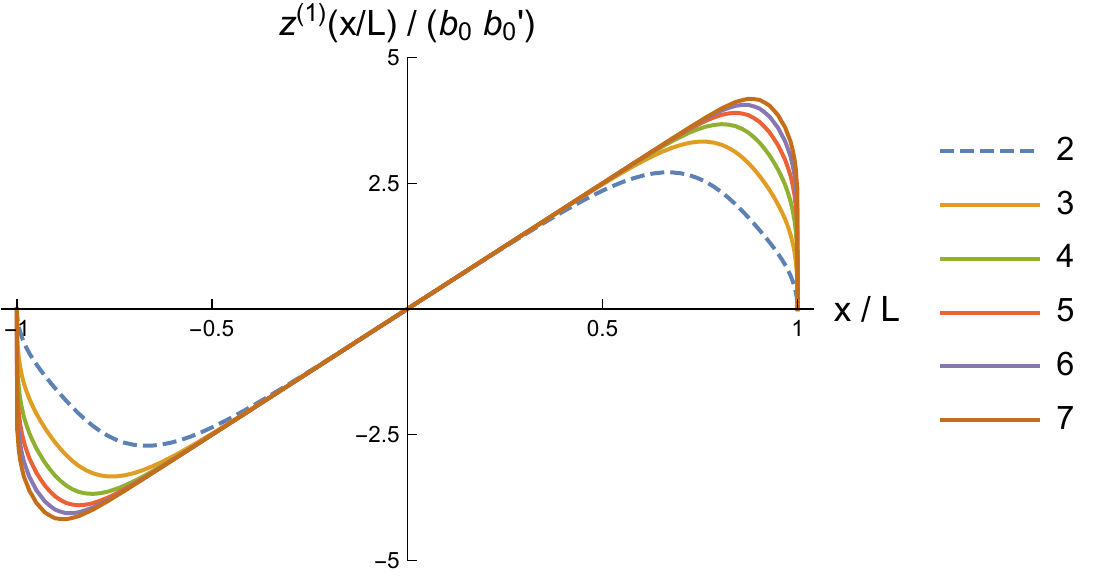}}
\caption{ Numerical solutions for $z^{(1)}( x/L)/(b_0\, b_0')$ for slabs of width $2L$ on a single boundary with $2 \le d \le 7$.  Away from the end points, increasing $L$ causes $z^{(1)}( x/L)$ to approach the first order correction to the horizon location.  Since $g^{(1)}_{AB}$ vanishes identically, this correction comes only from expanding $b=b_0+b_0'x+\dots$ within the zeroth order ansatz.
This correction is thus linear in $x$, given in this approximation by $z_H^{(1)}( x)= b_0' \,  x$.
}
\label{surf1}
\end{figure}

\begin{figure}[h!]
\centering
\subfloat[]{\includegraphics[width=0.45\textwidth]{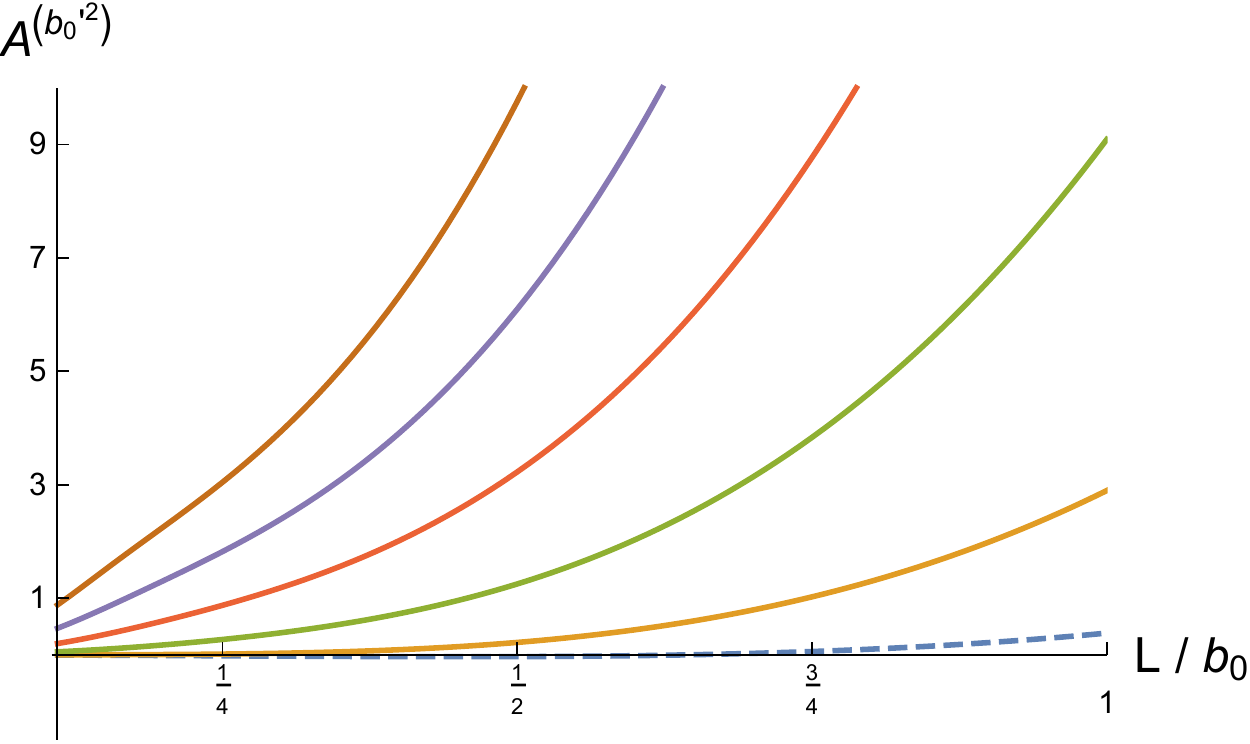}}\qquad
\subfloat[]{\includegraphics[width=0.45\textwidth]{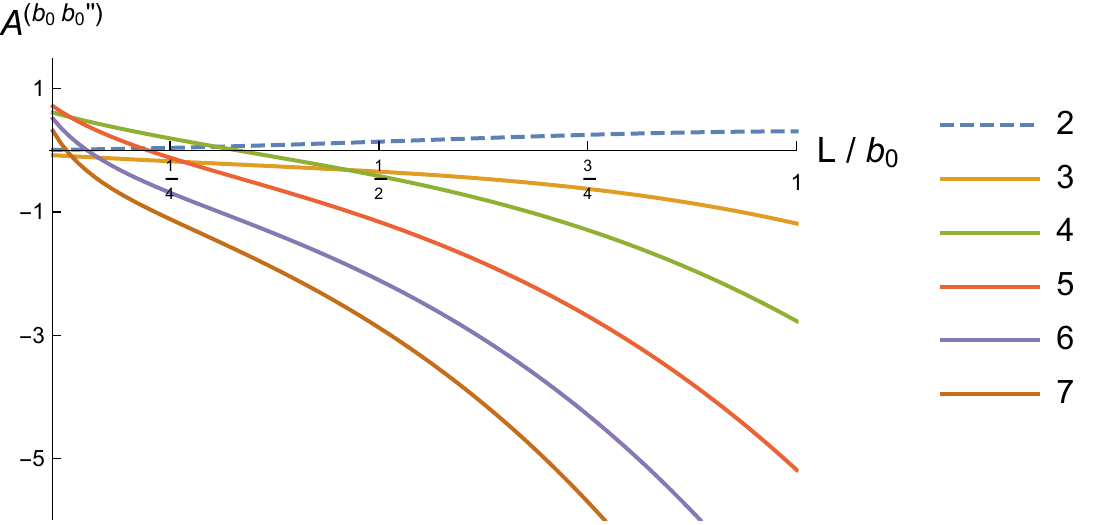}}
\caption{ Plots of (a) ${\sf A}^{(b_0'^2)}(L/b_0)$ and { (b) ${\sf A}^{(b_0b_0'')}(L/b_0)$ }for slabs of width $2L$ on a single boundary with $2\leq d \leq 7$.
}
\label{A2}
\end{figure}

Computing the second order change in area \eqref{genO2} requires only knowledge of the surface to $O(\epsilon)$.  This is because the order-zero surface is minimal, so changes in the area computed with with zeroth order metric are quadratic in changes of the surface.  The first order equation of motion is complicated, but is straightforward to work out and can be solved numerically by the same techniques as at order $\epsilon^0$. Results for $z^{(1)}(x/L)/(b_0\,b_0')$
are shown in figure \ref{surf1} for various dimensions and interval sizes.  The second order correction to the area then follows by summing the following three contributions: the above-mentioned quadratic change in the area computed using the zeroth-order metric due to the shift in the minimal surface at $O(\epsilon)$, the change in the area of the zeroth-order minimal surface due to the inclusion of $O(\epsilon^2)$ terms in the metric, and a cross-term linear in both the $O(\epsilon)$ shift of the surface and the $O(\epsilon)$ correction to the metric. In terms of the densitized area $\mathcal A_{\text{ren}}$, this correction takes the form
\ban{
A^{(2)}_\text{ren} = &\frac 12 \int dx  \left[\left. \left(\frac{\partial^2 \mathcal A_{\text{ren}}^{(0)}}{\partial z^2}- \deriv{}x\left(\frac{\partial^2  \mathcal A_{\text{ren}}^{(0)} }{\partial z \, \partial z'}\right)\right)\right|_{z^{(0)}(x)} \left(z^{(1)}(x)\right)^2+ \left. \frac{\partial^2\mathcal A_{\text{ren}}^{(0)}}{{\partial z'}^2}\right|_{z^{(0)}(x)} \left({z^{(1)}}'(x)\right)^2   \right]\mathbreak
&+ \int dx \left[\left. \left( \pd {\mathcal A_{\text{ren}}^{(1)}}{z} - \deriv{}x \left( \pd{\mathcal A_{\text{ren}}^{(1)}}{z'}\right)\right) \right|_{z^{(0)}(x)} z^{(1)}(x)\right]\mathbreak
&+\int dx \left . \mathcal A_{\text{ren}}^{(2)}\right|_{z^{(0)}(x)} \, , \label{generalshift}
}
where each line corresponds to one of the above three contributions described above. Numerical results are shown in figure \ref{A2}.

As a check on our numerics, we now compare with the analytic expressions of section \eqref{ABTZanalytics}.  We first consider the case $b/(\epsilon \, b') \gg L \gg b_0$.  At order $\epsilon^0$ we numerically compute $b_0\,{\sf A^{(0)}}/ L$ for large $L/b_0$, while at order $\epsilon^2$ we compute $b^3_0\,{\sf A^{(2)}}/ L^3$.  Results are tabulated in figure \ref{largeL} which shows agreement with \eqref{largeL0}.

\begin{figure}[h!]
\centering
\begin{tabular}{c ||c |c || c | c || c | c}
$d$ & $b_0 \, {\sf A}^{(0)} /L $ &Pred. & $b_0^3\,  {\sf A}^{(b_0'{}^2)} /L^3 $ & Pred. & $-b_0^3\,  {\sf A}^{({b_0 b_0''})} /{L^3} $ &Pred.\\
\hline
$2$ & 4.000 & 4.000 & 1.33 & 1.33 & 0.667 & 0.667\\
$3$ & 5.04 & 5.04& 5.04 & 5.04 &1.68 & 1.68\\
$4$ & $5.66$ & 5.66& 11.3  & 11.3 & $2.83$ & 2.83\\
$5$ &$ 6.06$ & 6.06& 20.2  & 20.2 & $4.04$ &4.04\\
$6$ & $6.35$ & 6.35& $31.7 $  & 31.7 & $5.29$ & 5.29\\
$7$ & $6.56$ & 6.56& $45.9 $ & 45.9 & $6.56$ & 6.56
\end{tabular}

\caption{Comparison of the numerically computed $L \gg b_0$ scaling of ${\sf A}(L/b_0)$ (left colums) from figure \ref{A2} with the predictions (Pred., right columns)  from \eqref{largeL0} for $2 \leq d \leq 7$. The numerical precision is at least three significant figures, estimated by comparing results for 100 and 150 lattice points and for fitting intervals $L/b_0 \in [40,50]$ and $L/b_0 \in [50,60]$}
\label{largeL}
\end{figure}

Turning now to the case $L \ll b_0$, we have verified that the coefficient of ${\sf A}^{(2)}$ proportional to ${b'_0}^2$ vanishes quadratically as $L \ll b_0$, and we may also numerically compute the $b_0\,b''_0$ contribution to $ \lim_{L \rightarrow 0} L^{d-4} {\sf A^{(2)}}$.  Our results are tabulated in figure \ref{small} and shown to agree with the analytic results \eqref{smallL} and \eqref{smallL4}.
\begin{figure}[h!]
\centering
\begin{tabular}{c ||c |c || c | c }
$d$ & $L^{d-2}\,{\sf A}^{(0)}  $ &Pred. & $ {L^{d-4}}\, {\sf A}^{({b_0 b_0''})}$ &Pred.\\
\hline
$3$ & $-0.718 $&$ -0.718$&$-0.729$&$ -0.729$\\
$4$ &$ -0.0802 $&$ -0.0802 $&$- 0.334 \, \log L$ & $ - 0.333 \, \log L$\\
$5$ &$ -0.00864\pm 0.00001$ &$ -0.00865 $&$ 0.0897\pm 0.0020 $&$0.0916$\\
$6$ &$ -0.000821 \pm 0.000002$&$ -0.000822 $&$ 0.00850\pm 0.00039 $&$ 0.00885$\\
$7$ &$ -0.0000684 \pm 0.0000002$&$ -0.0000685 $&$ 0.000834 \pm 0.000041 $&$ 0.000871$
\end{tabular}

\caption{Comparison of the numerically computed $L \ll b_0$ scaling of ${\sf A(L/b_0)}$ (left colums) from figure \ref{A2} with the predictions (Pred., right columns)  of \eqref{smallL}, \eqref{smallL4} for $3 \leq d \leq 7$. The numerical precision (estimated as in figure \ref{largeL}) is shown when it falls below three significant figures.}
\label{small}
\end{figure}

\subsection{Entropy for pairs of diametrically opposed slabs}
\label{ATFDopposed}

We now consider the entropy of a pair of corresponding slabs on opposing boundaries. Both slabs are defined by $|x-x_0| \le L$,  but one lies at $\theta=0$ while the the other lies at $\theta = \pi$. Without loss of generality we again set $x_0=0$ in this section.  As in \cite{Morrison:2012iz,Hartman:2013qma}, for $L \ll b$ the minimal surface will be simply two copies of the one found in section \ref{ATFDsingle}, so that the mutual information between these two slabs vanishes.  But for $L \gg b$ the minimal surface represents a different phase, again having two disconnected pieces but now with each localized near $x = \pm L$.    Here the slabs share non-zero mutual information $I$.  In this phase the entropy is independent of $L$ and depends only on the local behavior of $b(x)$ near $x = \pm L$.    Note that the contribution from each surface is just the entropy one would compute for a pair of half-spaces, both defined by  $x > L$ (or  $x < - L$) but on opposite boundaries. For simplicity we thus focus on this `half-TFD' entropy below. All quantities associated with the half-TFD problem will be marked with hats ($\hat{}$) to distinguish them from the corresponding quantities of section \ref{ATFDsingle}.

As before, computing the area to order $\epsilon^2$ requires only knowledge of the entangling surface to first order.  It thus suffices to write
\ban{
\hat x(z) =\hat x^{(0)}(z) + \epsilon \,\hat x^{(1)}(z) + \cdots,
}
At zeroth order the entangling surface relevant to this half-TFD problem lies at precisely $\hat x^{(0)}(z)=\pm L$ and extends from one boundary to the other, passing through to the horizon. The total area at this order may be computed analytically and we find
\ban{
{\hat A}^{(0)}_\text{ren} =& V\, \ell^{d-1} \frac{2^{1-4/d }}{b^{d-2}}\left( \frac{d -4}{d-2}\right)\left(2-16^{1/d } \, _2F_1\left[\frac{2}{d }-1,\frac{4}{d };\frac{2}{d };-1\right]\right) \ .
\label{CB0}
}

\begin{figure}[h!]
\centering
\includegraphics[width=0.65\textwidth]{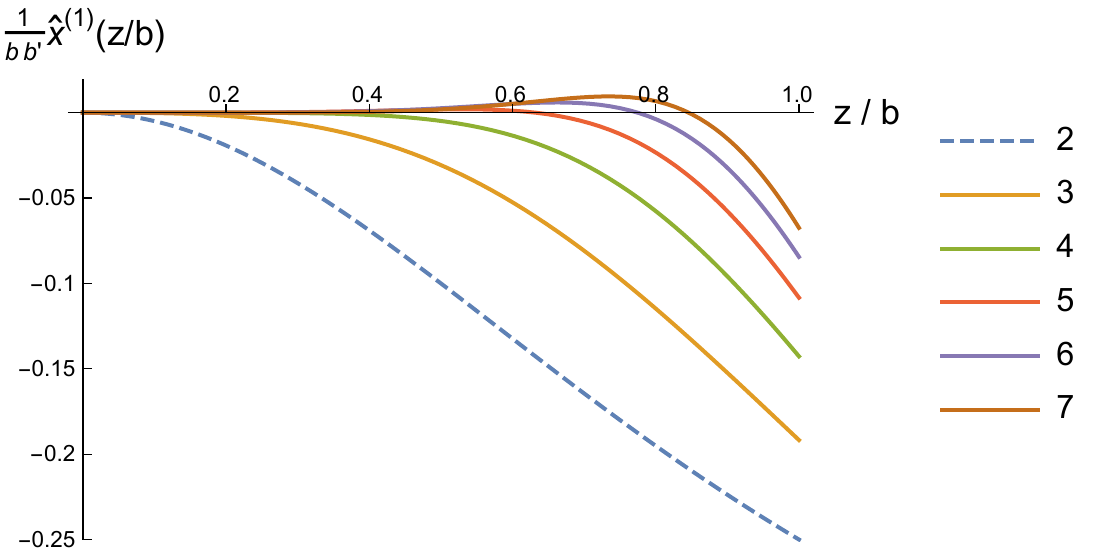}
\caption{ Numerical solutions $\frac{\hat x^{(1)}(z/b)}{bb'}$ for the half-TFD problem with $2\leq d \leq 7$, with $bb'$ evaluated at $ x = \pm L$. In the large $d$ limit, one may show analytically that this function vanishes everywhere except at the horizon.
}
\label{crosssurf}
\end{figure}

At first order we proceed numerically, with $\hat x^{(1)}(z)$ satisfying the first order equation of motion
\ban{
0&= \left(b^d+z^d\right) \left(b^d+z^d\right)^{4/d} \partial_z^2{\hat x^{(1)}}+\frac 1z  \left((d+1) z^d-(d-1) b^d\right) \left(z^d +b^d\right)^{4/d} \partial_z \hat x^{(1)}+2 b^4 (d-2) b' z^d.
}
We simplify the analysis by using the symmetry that relates our two boundaries.  We thus compute the area for a surface extending from one boundary to the horizon and multiply by 2.  The boundary conditions are that $\hat x=\pm L$ at $z=0$ and that $\frac{d\hat x}{dR}=0$ at the horizon $R=0$, where $R$ is the regular coordinate associated with \eqref{regular}.  But since it is convenient to work in terms of the original $z$ coordinate, we note that to order $\epsilon$ this is equivalent to imposing the boundary condition
\ban{
\left. \hat x'(z)\right|_{z=b} = - \frac \epsilon {16} b\, b' \, . \label{bcCB}
}
We solve numerically for the minimal surface in the region $z\in [0,b]$ and simply approximate $\hat x(z)$ by $\hat x=\pm L$  in the order $\epsilon^2$-sized region $z \in [b,\tilde b]$.  Numerical solutions for $\hat x^{(1)}(z/b)$  are shown in figure \ref{crosssurf}  for $2 \le d \le 7$.

The second order area correction now has an additional contribution due to the $O(\epsilon^2)$ shift in the endpoint of the minimal surface. This contribution can be computed analytically and the full second order shift is given by
\ban{
{\hat A}^{(2)}_\text{ren}&=\tilde A^{(2)}_\text{ren}
+\epsilon^2\,V \, \ell^{d-1} \frac{2^{\frac{2d-8}{d}}}{d^3 b^{d-3}}{  \left(2\, d \left(\, _2F_1\left[1,-\frac{2}{d};\frac{2}{d};-1\right]-3\right)-\frac{2 \sqrt{\pi }\, \Gamma \left(\frac{2}{d}\right)}{\Gamma \left(\frac{1}{2}+\frac{2}{d}\right)}\right)}\left. \partial_z \gp{tt}\right|_{z=b}  \,  \label{CB2}
}
where $\tilde A^{(2)}_\text{ren}$ includes the area of only the part of the surface with $z\le b$.   Note that \eqref{CB2} depends on $L$ only through evaluating $b$ (and its derivatives) at $ x = \pm L$.  We compute \eqref{CB2} numerically.  Results are displayed in figure \ref{CB} in terms of dimensionless coefficients defined by
\begin{equation}
\label{posIcoeff}
{\hat A}^{(2)}_\text{ren}  = \frac{V \, \ell^{d-1}}{b^{d-2}} \hat{\sf A}^{(2)} \, , \text{ with} \ \
 \hat{\sf A}^{(2)} =  {b'}^2 \hat{\sf A}^{({b'}^2)} +  b\,b'' \hat{\sf A}^{(bb'')}\,,  d\neq 2, 4.
\end{equation}
Here $b, b',b''$ are evaluated at $ x=\pm L$.  For $d=2,4$ we use analogous notation but with logarithmic subtractions as in \eqref{d24}.
\begin{figure}[h!]
\centering
\begin{tabular}{c ||c |c}
$d$ & $  \hat{\sf A}^{({{b'}^2})} $& $  \hat{\sf A}^{({ b\, b''})} $\\
\hline
$2$ & 0.00 & 0.00\\
$3$ & 0.531 & -0.294 \\
$4$ &0.571 & {{0.0716}} \\
$5$ & 0.815 & -0.142 \\
$6$ & 1.28 & -0.562 \\
$7$ & 1.93 & -0.845
\end{tabular}
\caption{The coefficients  $ \hat{\sf A}^{({b'}^2)} $ and $\hat{\sf A}^{(b\, b'')} $ for the half-TFD problem for $2 \leq d \leq 7$. The numerical precision is estimated by comparing results for $100$ and $150$ lattice points, giving better than one part in $10^{-10}$.}
\label{CB}
\end{figure}

\subsection{Phase transition}
\label{ATFDphase}

We now analyze the transition between the $I=0$ and $I>0$ phases for a pair for $|x-x_0| \le L$ slabs on opposite boundaries.  In particular, we compute the effect of inhomogeneities on the critical length $L_\text{crit}$.

For this purpose, we should compare twice the area of the entangling surface for a slab $|x|\leq L$ on a single boundary with that of the sum of the surfaces for the half-TFD problems at $x=\pm L$.  The phase transition will occur when $L$ is of order $b$, so at small $\epsilon$ we have $L \ll b/(\epsilon b')$ and we may expand $b(\pm L)$ in \eqref{posIcoeff} in a Taylor series.  At order $\epsilon^0$, the surfaces at $x = \pm L$ have equal area, so we can determine the zeroth order value of $L_\text{crit}$ by comparing (twice) the numerical value of \reef{CB0} for $b=b_0$ with (twice what is shown in) figure \ref{A0}. Results are displayed in figure \ref{lcrit}.

As discussed in section \ref{ATFDsingle}, the first order correction to the area of the connected surface vanishes. For the disconnected surfaces, we have a first order correction from expanding \reef{posIcoeff}.  But this correction is proportional to $ x \,b'_0$, so the corresponding contributions cancel between the surfaces at $x = \pm L$; there can be no change in $\lc$ at first order.

\begin{figure}[h!]
\centering
\subfloat[]{\includegraphics[width=0.5\textwidth]{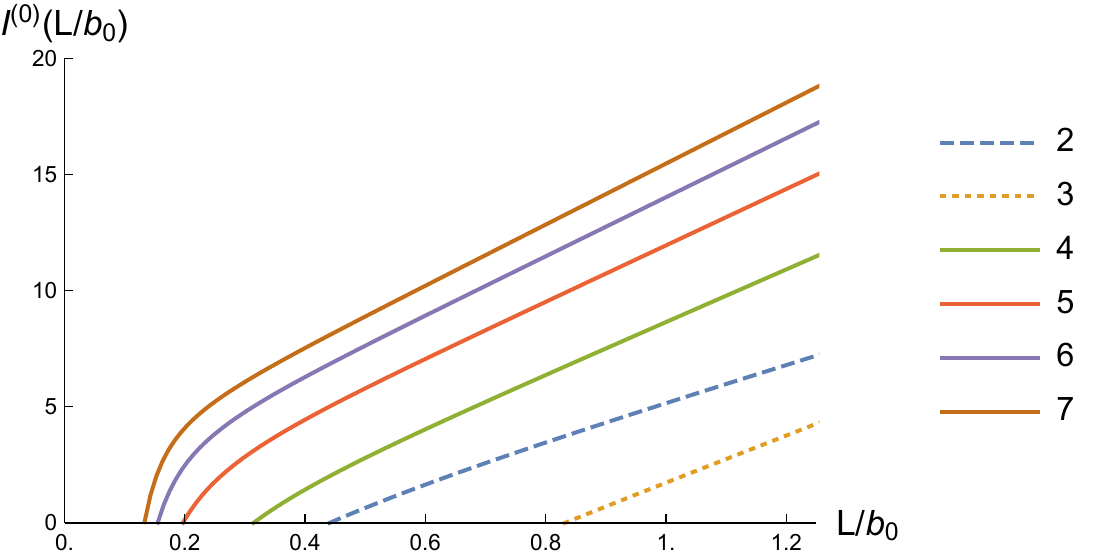}}\\
\subfloat[]{\includegraphics[width=0.435\textwidth]{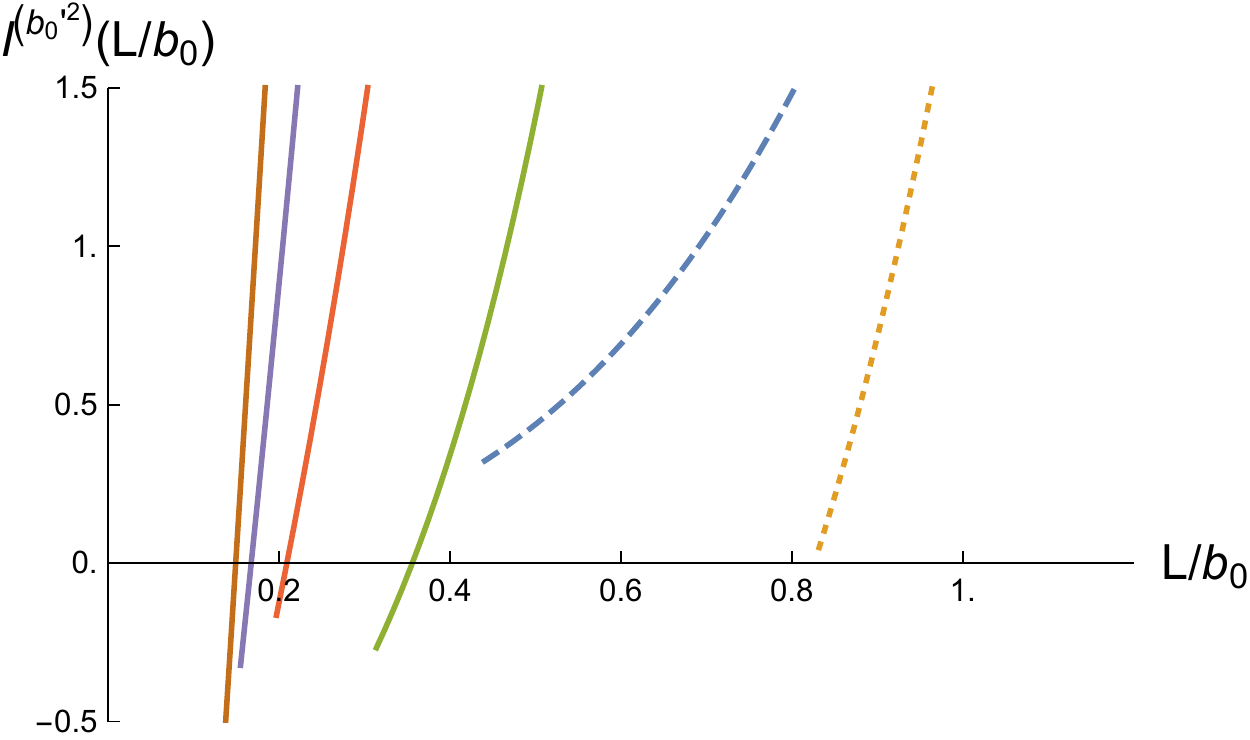}} \hspace{1cm} \subfloat[]{\includegraphics[width=0.48\textwidth]{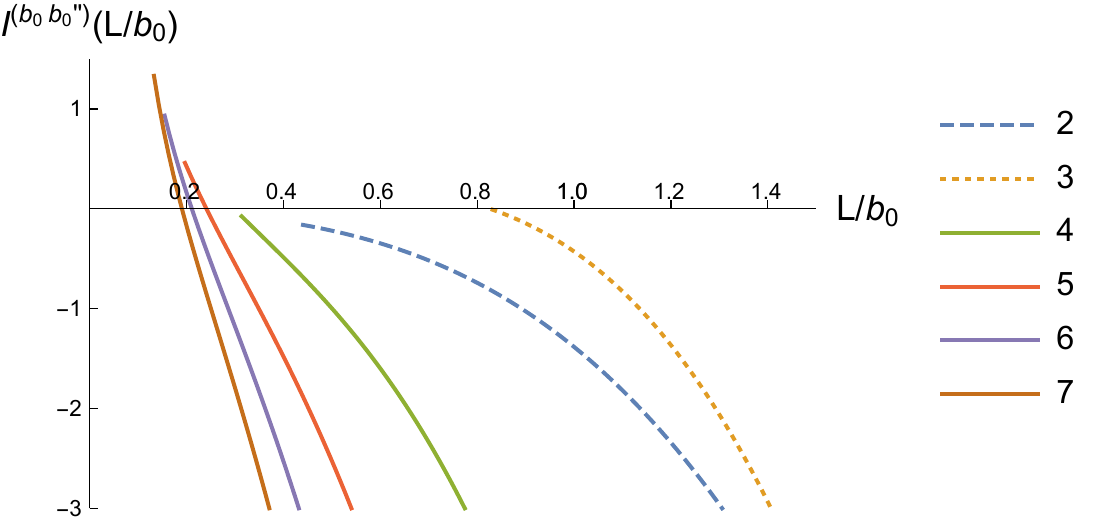}}
\caption{
The coefficients of ${\sf I}(L/b_0)$ for $2 \leq d \leq 7$ to second order. The mutual information vanishes for $L<L_\text{crit}$.
}
\label{MI}
\end{figure}

At second order, we can write $\lc = \lc^{(0)} + \epsilon^2 \lc^{(2)}$ and solve \begin{equation}
\label{ATFDtrans}
2A_\text{ren}(\lc) =\hat A_\text{ren}|_{x=L_c} + \hat A_\text{ren}|_{x=-L_c}.
 \end{equation}
Here it is useful to note that Taylor expanding $\hat A_\text{ren}|_{x=\pm L_c}$ and then performing our adiabatic expansion gives
\begin{eqnarray}
\hat A_\text{ren}|_{x=L_c} + \hat A_\text{ren}|_{x=-L_c}  &=& 2 \hat A_\text{ren}|_{x=0} + \lc^2 \partial_x^2 \hat A_\text{ren}|_{x=0} + \dots \cr &=& 2 \hat A^{(0)}_{ren}|_{x=0} + \epsilon^2 \left( 2 \hat A^{(2)}_{ren}|_{x=0}  + \left(\lc^{(0)}\right)^2 \left[ (b_0')^2 \partial_b^2 \hat A^{(0)}_{ren}|_{x=0} + b_0'' \partial_b \hat A^{(0)}_{ren}|_{x=0} \right]  \right) \cr &&+ O(\epsilon^4).
\end{eqnarray}
Solving \eqref{ATFDtrans} to order $\epsilon^2$ then gives
\ban{
\label{L2}
\lc^{(2)} &=  \frac{   \frac{\left(\lc^{(0)}\right)^2}{2} \left( (b_0')^2 \partial_b^2 \hat A^{(0)}_{ren}|_{x=0} + b_0'' \partial_b \hat A^{(0)}_{ren}|_{x=0} \right) + \hat A^{(2)}_{ren}|_{x=0}   - A^{(2)}(\lc^{(0)})\,
} {\left.\partial_L A_{ren}^{(0)}(L/b_0)\right|_{\lc^{(0)}}}
.}
Figure \ref{lcrit} displays numerical results for $2\leq d \leq 7$  in terms of the coefficients defined by
\ban{
\lc^{(2)} /b_0 = {b_0'}^2\,{\sf L}^{({b_0'}^2)}+{b_0 b_0''}\,{\sf L}^{({b_0 b_0''})}\, .
\label{lcritcoeff}
}
\begin{figure}[h!]
\centering
\begin{tabular}{c||c||c |c }
$d$ & $L^{(0)}_\text{crit}/ b_0$& $   {\sf L}^{({b'_0}^2)} $& $ {\sf L}^{(b_0b''_0)}$\\
\hline
$2$&$ 0.441 $&$ -0.0285 $&$ 0.0143 $\\
$3$&$ 0.832 $&$ -0.00532\pm 0.00027 $&$ 0.00111\pm 0.00012 $\\
$4$&$ 0.314 $&$0.0132 \pm 0.0004$&$ 0.00417 \pm 0.00021$\\
$5$&$ 0.197 $&$ 0.00305  \pm 0.00048 $&$ -0.00983\pm 0.00024$\\
$6$&$ 0.155 $&$ 0.00300 \pm 0.00057$&$  -0.0104\pm 0.0002$\\
$7$&$ 0.133 $&$ 0.00405 \pm 0.00090 $&$ -0.00912\pm 0.00032  $\\
$8$&$ 0.119 $&$ 0.00872 \pm 0.0015 $&$ -0.00834\pm 0.00045  $
\end{tabular}
\caption{The coefficients governing $\lc$ to second order. The numerical precision is shown when it falls below three figures when estimated as described in appendix \ref{error}. The numerical result for $d=2$ (shown) agrees with analytic predictions from appendix \ref{2+1}.}
\label{lcrit}
\end{figure}
In addition, figure \ref{MI} shows result for the mutual information between the slabs using the notation
\ban{
\hat {I} &= \frac{V\, \ell^{d-1}}{b_0^{d-2}} \hat{\sf I}(L/b_0)\, ,  \mathbreak
 \hat {\sf I}(L/b_0) & = \hat{\sf I}^{(0)}(L/b_0)+ \epsilon^2\left( b_0'{}^2\, \hat{\sf I}^{(b_0'{}^2)}(L/b_0)+ b_0b_0''\, \hat{\sf I}^{(b_0 b_0'')}(L/b_0)\right) \, .
}

We find to second order that $\hat I$ has an interesting dependence on dimension. First although $\hat{\sf I}^{(b_0'{}^2)}$ is positive for most $L > L_{crit}$, for $d\geq 4$ it becomes slightly negative near $L_{crit}$.  As a result, a non-zero $b_0'$ causes $L_{crit}$ to increase for $d\geq4$ and decrease for $d=2,3$.  The effect of second derivatives depends on dimension as well: a positive $b_0''$ increases $L_\text{crit}$ for $2\leq d \leq 4$ but decreases $L_\text{crit}$ for $5\leq d \leq 7$.  For $d=2$ the above behavior is derived analytically in appendix \ref{2+1}; it would be interesting to develop an analytic understanding of the higher dimensional results as well. Due to the many interesting features in this data, we take extra care to understand the convergence of our numerics in appendix B.

%%%%%%%%%%%%%%%%%%%%%%%%%%%%%%%%%%%%%%%%%%%%%%%%%%%%%%%%%%%%%%%%%%%%%%%%%%%%%%%%
%%%%%%%%%%%%%%%%%%%%%%%%%%%%%%%%%%%%%%%%%%%%%%%%%%%%%%%%%%%%%%%%%%%%%%%%%%%%%%%%
%%%%%%%%%%%%%%%%%%%%%%%%%%%%%%%%%%%%%%%%%%%%%%%%%%%%%%%%%%%%%%%%%%%%%%%%%%%%%%%%

\section{States of Confining Theories}
\label{confined}

We now turn to the second interpretation in which our path integral computes the ground state of a confining gauge theories on the surface $y^1=0$.  This necessarily restricts our discussion to $d \ge 3$.

We again consider slabs $|x-x_0| \le L$.  As in section \ref{ATFDopposed}, there are two possible phases for the minimal surface. For $L \ll b$ the minimal surface is connected and does not reach $R=0$.  But there is also another local extremum of the area given by a disconnected surface that consists of two disks, each localized near $x-x_0 = \pm L$.  At small $L$ the disconnected surface has larger area, though increasing $L$ leads to a phase transition at which the disconnected surface becomes minimal.  Interestingly, at still larger values of $L$ the connected extremum becomes singular and ceases to exist.  The two phases are
shown in figure \ref{phasetrans} and will be studied in sections \ref{conf narrow} and \ref{confwide}  below.

The general feature that the entanglement becomes independent of $L$ at large $L$ is to be expected in confining theories, as they have finite correlation lengths.  But the sharp phase transition seen here is a feature of large $N$ \cite{Klebanov:2007ws,Nishioka:2006gr}. \begin{figure}[h!]
\centering
\subfloat[]{\includegraphics[width=0.4\textwidth]{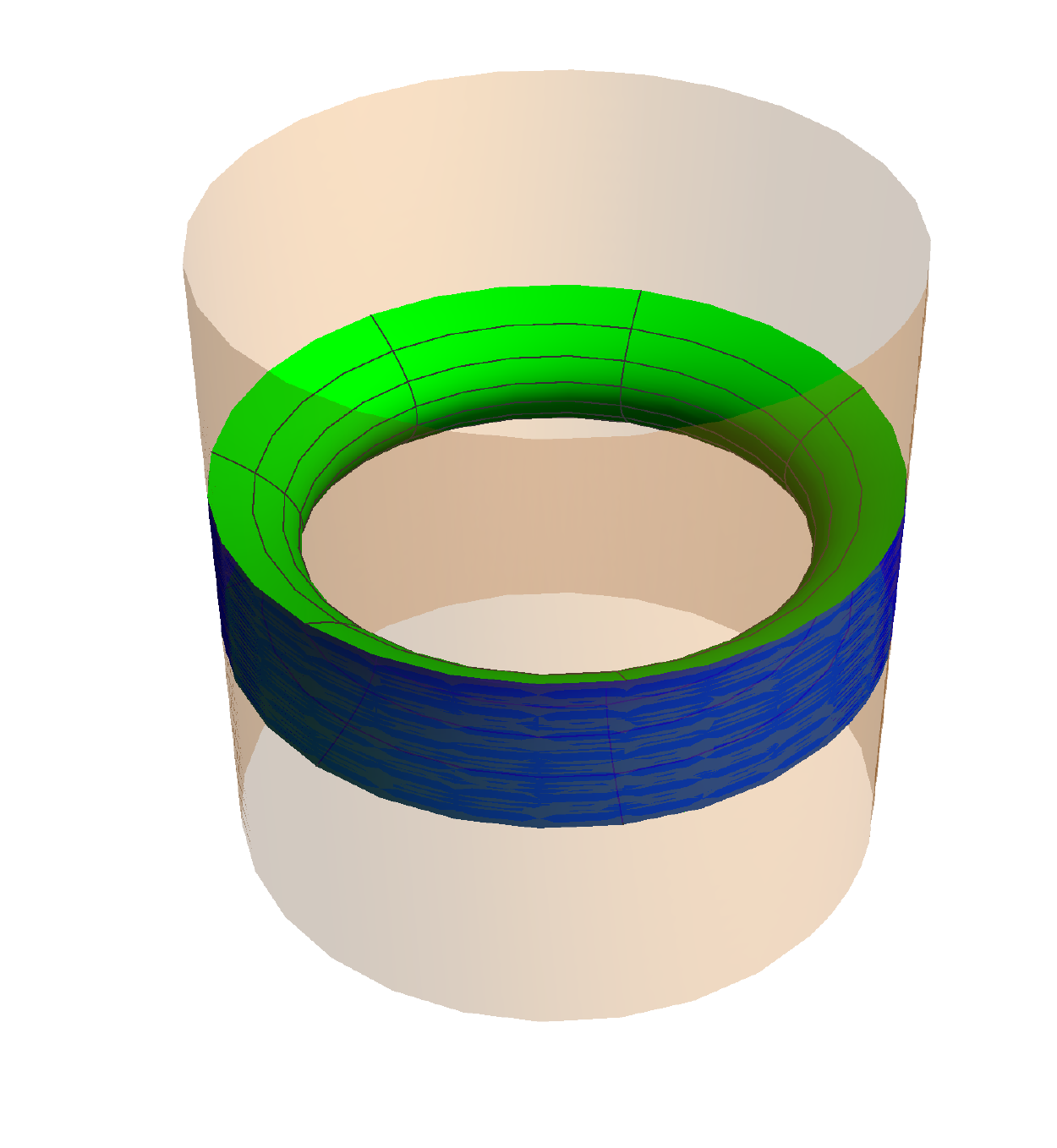}}\qquad
\subfloat[]{\includegraphics[width=0.4\textwidth]{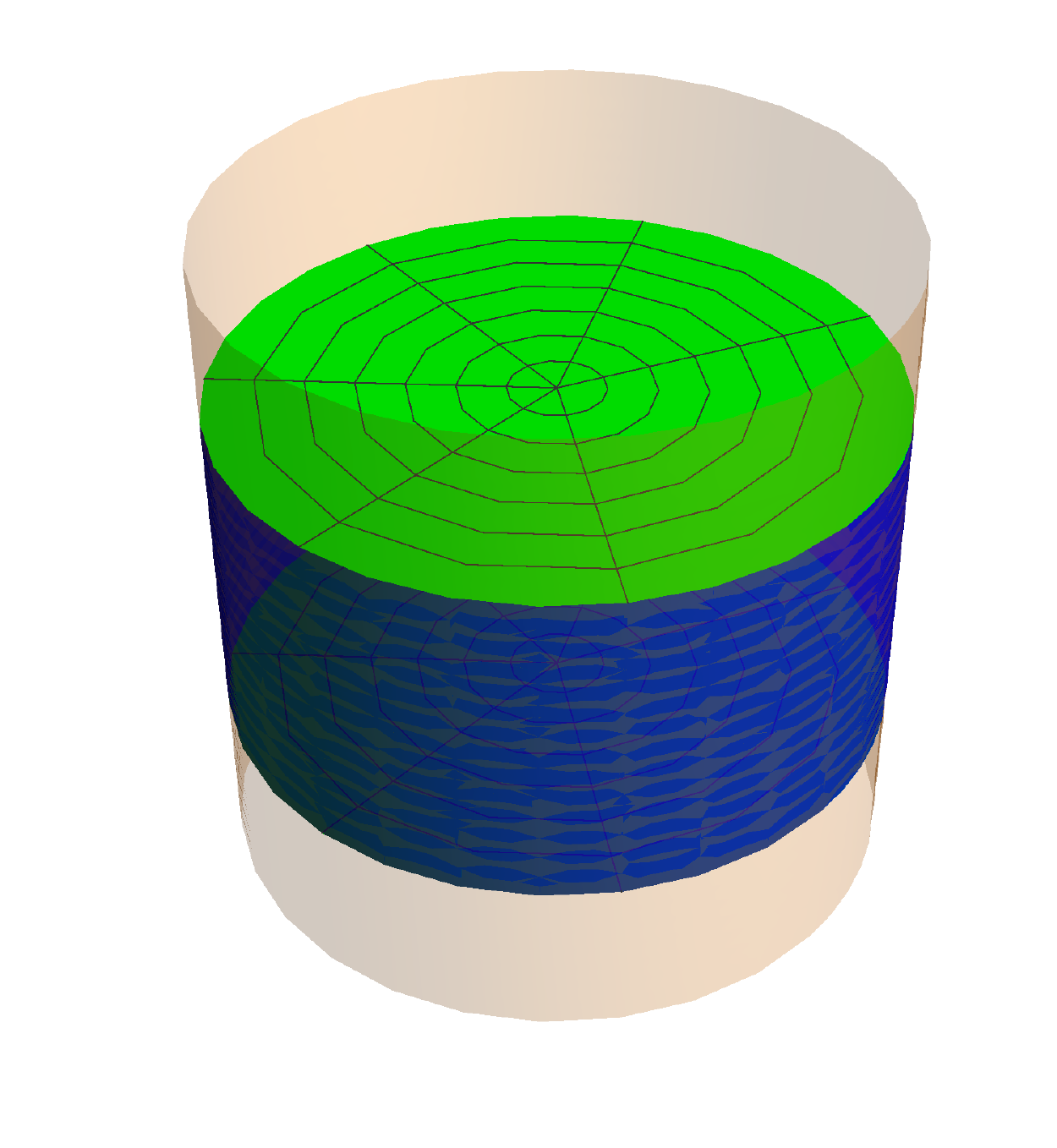}}
\caption{Possible topologies for the extremal surfaces for a strip on the boundary. As shown in (a), for thinner strips the connected surface has minimal area. For thicker strips, the disconnected surface consisting of two disks shown in (b) has minimal area.
}
\label{phasetrans}
\end{figure}

Below, we find it useful to write
\ban{
A_\text{bare} &= 2\pi\,  \tilde V\, \ell^{d-1} \int \mathcal A_\text{bare}\, d\lambda \ \ \text{for} \mathbreak
\mathcal A_\text{bare}&=  g_{yy}^{(d-3)/2} g_{\theta\theta}^{1/2} \left(\frac{z'(\lambda)^2}{z(\lambda)^2} +  x'(\lambda)^2\, g_{xx}\right)^{1/2}, \label{confunc}
}
where $\tilde V$ is the volume of a $(d-3)$ torus that we use to regularize the $y^2,\dots, y^{d-2}$ directions. To compute the entropy, we must as usual find the minimal surface to $O(\epsilon)$. We will also need the explicit counterterms that renormalizing the area functional to second order. The computations are analogous to those in section \ref{deconfined}, though now the minimal surface equations lead to the  asymptotic expansion
\ban{
x(z) &=  x_B + \frac {\epsilon\,  b'} {2\,(d-2)}\,z^2 + c_d \, z^d + O(z^{d+1},\epsilon^2) \ ,
\label{confexpx}}
where $x_B$ is the point of intersection with the boundary.  Inserting \eqref{confexpx} into \reef{confunc} gives
\ban{
\mathcal A_\text{bare} &=\frac { \alpha_d \,b}{z^{d-1}}-\epsilon^2\, \frac{\alpha_d (d-3)}{2(d-2)^2}\, \frac {{b'}^2}b\, \frac 1{z^{d-3}}+ \epsilon^2 \, \frac {\alpha_d b''}{2 (d-1)(d-2)} \frac 1 {z^{d-3}}+O(z^0),
\label{confbareexp}
}
so for $d> 4$ we may take
\ban{
A_\text{ct} &=
 2\pi \tilde V\, \alpha_d\,\ell^{d-1}\left[- \frac 1 {(d-2)} \frac b {z^{d-2}}+\epsilon^2\, \frac{(d-3)}{2(d-2)^2(d-4)}\, \frac {{b'}^2}b\, \frac 1{z^{d-4}}- \epsilon^2 \, \frac {b'' }{2 (d-1)(d-2)(d-4)} \frac 1 {z^{d-4}}\right].
\label{ct}
}
In lower dimensions we have
\ban{
A_\text{ct}&= 2\pi\, \alpha_d \tilde V \, \ell^3 \left[-\frac 12 \frac b{z^2}+\epsilon^2\,\left(\frac {b''}{12} - \frac 18  \frac {{b'}^2}b\right) \log (z/\ell)+{ \epsilon^2 \left(-\frac {b'' }{24} + \frac 18\frac {{b'}^2}b \right)}\right]  \hspace{0.75cm} d=4\mathbreak
A_\text{ct} &= -2\pi \,\alpha_d \, \ell^2 \frac b {z}\hspace{4.85cm} d=3  \, ,
}
{ where  the counterterms again match the covariant prescription of \cite{Taylor:2016aoi}, whose details we have again used to fix the $z$-independent terms for $d=4$.}
We can now compute the area of the minimal surface for the regimes $L\ll b$ and $L\gg b$ and study the phase transition between connected and disconnected topologies. Additionally, without loss of generality we set $x_0=0$ for the rest of this section.

\subsection{Narrow slabs}
\label{conf narrow}

We begin with the regime $L \ll b$, where the entropy will be given by the connected surface \cite{Klebanov:2007ws,Nishioka:2006gr}.  The computations proceed much as in section \ref{ATFDsingle}, though  we are no longer able to obtain analytic results for the second order area in the large and small $L$ limits.  Indeed, this phase fails to exist at sufficiently large $L$, while for the small $L$ limit
the first order correction $z^{(1)}(x)$ must be computed numerically
even in the approximate geometry \reef{FGapprox}. However, the expansion \eqref{FGapprox} does require the leading small $L$ behavior of $A_{\text{ren}}^{(2)}$ to be of order $L^{4-d}$. As a test of our numerics, we compare below the coefficient of $L^{4-d}$ computed using the full metric against that computed using the truncated metric \eqref{FGapprox}. At zeroth order we can compare against an analytic prediction, as at this order \reef{FGapprox} is just Poincar\'e AdS${}_{d+1}$ and $\theta$ acts just like a $y$-coordinate with period $2\pi \alpha_d b$. As a result, the area is given by \reef{A0FG} with $V= 2\pi \alpha_d\, b\,\tilde V$.

\begin{figure}[h!]
\centering
\subfloat[]{\includegraphics[width=0.5\textwidth]{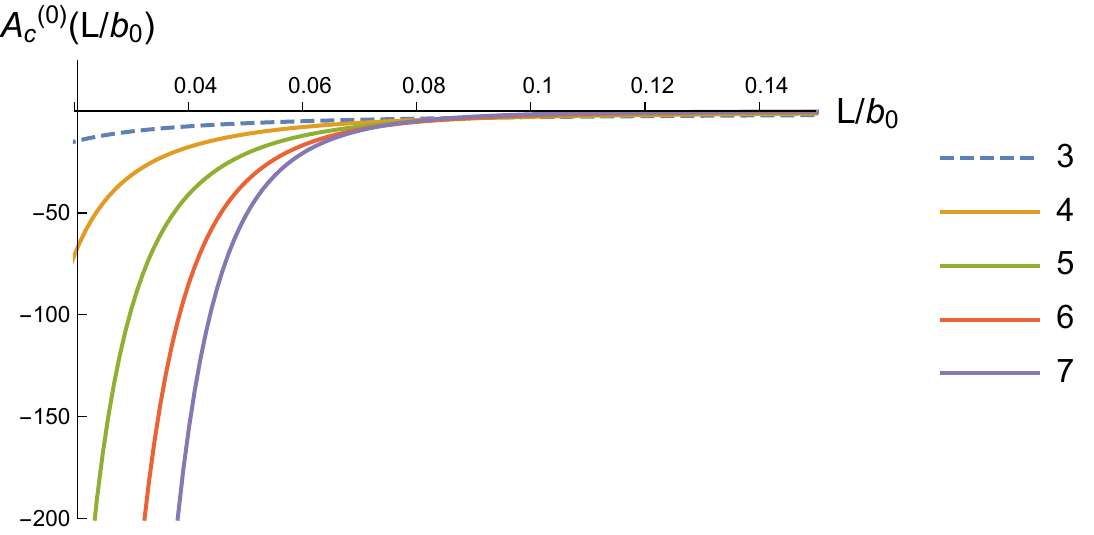}}\\
\subfloat[]{\includegraphics[width=0.45\textwidth]{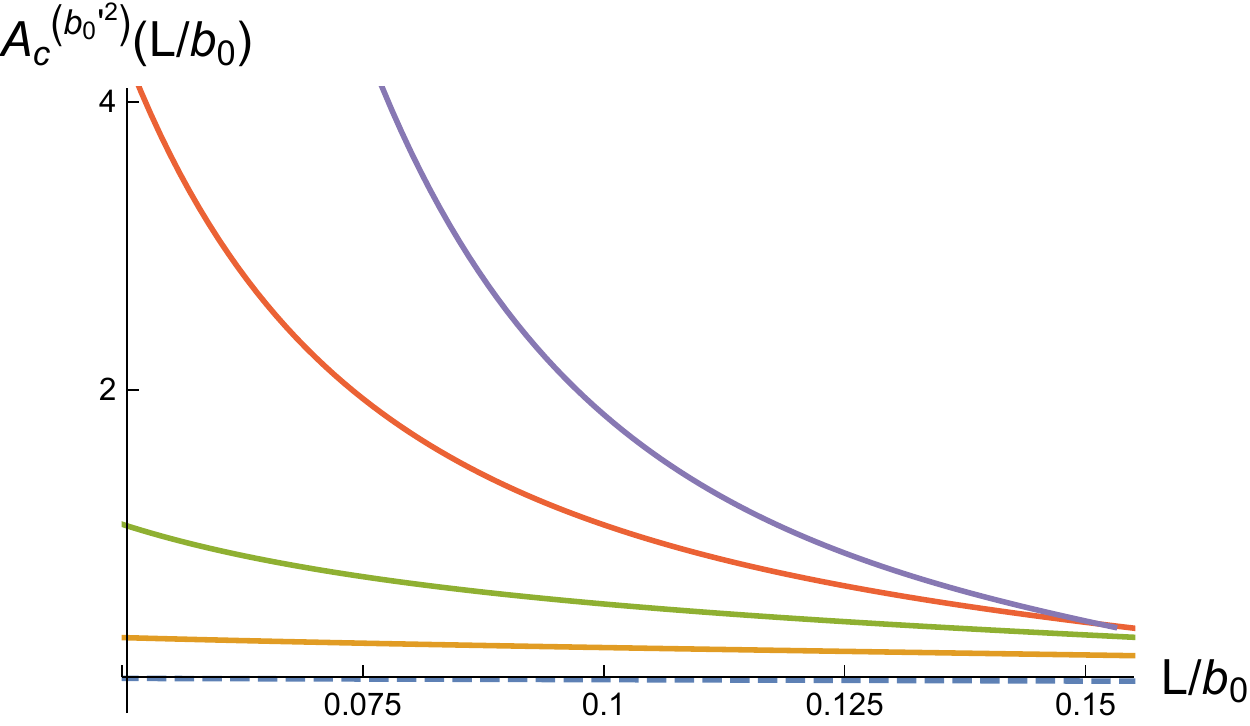}\qquad\includegraphics[width=0.5\textwidth]{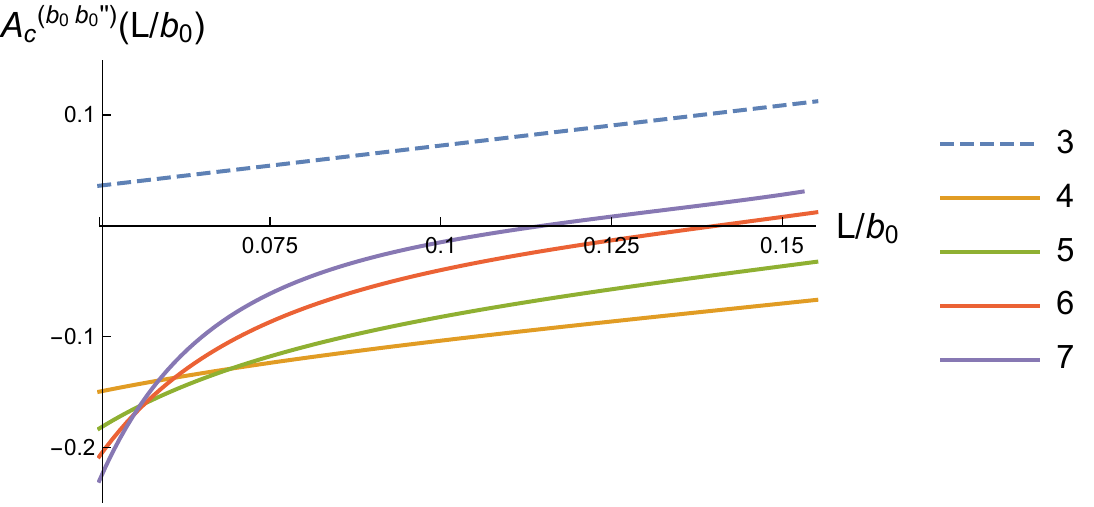}}
\caption{Numerical results for ${\sf A}^{(0)}_c(L/b_0)$,  ${\sf A}^{(b'_0{}^2)}_c(L/b_0)$, and ${\sf A}^{(b_0b''_0{})}_c(L/b_0)$ for $3 \leq d \leq 7$.
 }
\label{conn}
\end{figure}

As in section \ref{ATFDsingle}, we consider the case $L\ll b/(\epsilon\, b')$ so to order $\epsilon^2$ the inhomogeneities are described by $b_0$, $b_0'$, and $b_0''$. We state our numerical results for the connected area in terms of the dimensionless function ${\sf A_c}(L/b_0)$ defined for $d\neq 4$ by
\ban{
A_\text{ren} =  \frac{2\pi \tilde V\,\ell^{d-1}}{ b_0^{d-3}} {\sf A_c}(L/b_0) \, \label{genC}.
}
where the subscript ${\sf c}$ will denote quantities associated with the connected entangling surface.
For $d=4$ it is useful to explicitly remove the $\log(\ell)$ dependence introduced by our regularization scheme.  We therefore write
\ban{
A_\text{ren} =  \frac{2\pi \tilde V\,\ell^{3}}{ b_0} {\sf A_c}(L/b_0)+ \epsilon^2\, {2\pi \alpha_d \tilde V\,\ell^{3}} \left( \frac{b_0''}{12} - \frac 18 \frac{{b_0'}^2}{b_0} \right) \log (\ell/b_0) \, .
}
As before, we use the adiabatic expansion to write
\ban{
{\sf A_c}(L/b_0)= {\sf A_c}^{(0)}(L/b_0)+ \epsilon \,{\sf A_c}^{(1)}(L/b_0)+ \epsilon^2 \,{\sf A_c}^{(2)}(L/b_0)+ O(\epsilon^3) \mathbreak
 \text{with}\hspace{0.25cm}  {\sf A_c}^{(2)}(L/b_0) = (b'_0)^2 {\sf A}^{(b'_0{}^2)}_{\sf c}(L/b_0) +  b_0\,b''_0 {\sf A}^{(b_0\,b''_0)}_{\sf c}(L/b_0)\,, \label{genC2}
}
where symmetry under $x \rightarrow -x$ again requires the first order correction to vanish.  Numerical results are displayed in figure \ref{conn}.

As a check on our numerics, we extract $\lim_{L \to 0} L^{d-2} A_\text{ren}^{(0)}$ and $\lim_{L \to 0} L^{d-4} A_\text{ren}^{(2)}$ and compare in figure \ref{smallLS} with the same coefficients as  determined by approximating the metric to $O(z^2)$ in the Fefferman - Graham expansion \reef{FGapprox}.
\begin{figure}[h!]
\centering
\subfloat[]{
\begin{tabular}{c ||c|c }
$d$ & $\lim_{L\to 0}L^{d-2} {\sf A_c}^{(0)}$ &Pred. \\
\hline
$3$ & $-0.301$&$ -0.301$\\
$4$ &$ -0.0283 $&$ -0.0283 $\\
$5$ &$ -0.00262$ &$ -0.00262 $\\
$6$ &$ -0.000218 $&$ -0.000217$\\
$7$ &$ -0.0000161 $&$ -0.0000160$
\end{tabular}}\\
\subfloat[]{
\begin{tabular}{c ||c |c || c | c }
$d$ & $\lim_{L\to 0}L^{d-4} \partial_{(b_0')^2} {\sf A_c}^{(2)}$ &Approx.& $\lim_{L\to 0}L^{d-4} \partial_{b_0\, b_0''}{\sf A_c}^{(2)} $ &Approx.\\
\hline
$3$ & $-0.186 \pm 0.003 $&$ -0.186 \pm 0.003$&$0.740 \pm 0.003$&$0.740 \pm 0.003$\\
$4$ &$ -0.0828\pm 0.0006 $&$ -0.0828\pm 0.0006 $&$0.516\pm 0.0008$ & $0.516\pm 0.0008$\\
$5$ &$ 0.0678\pm 0.0044$ &$ 0.0678\pm 0.0044 $&$ -0.00948 \pm 0.00004$&$ -0.00949 \pm 0.00004$\\
$6$ &$ 0.0189 \pm 0.0027$&$ 0.0189 \pm 0.0027 $&$-(5.70\pm 0.08) \times 10^{-4}$&$-(5.70\pm 0.08) \times 10^{-4}$\\
$7$ &$ 0.00400 \pm 0.00077$&$ 0.00400 \pm 0.00077 $&$ -(3.49\pm 0.12) \times 10^{-5} $&$ -(3.49\pm 0.12) \times 10^{-5} $
\end{tabular}}\\

\caption{Comparison of the numerically computed $L\ll b_0$ scaling of $\sf{A}(L/b_0)$ for $3\leq d\leq 7$ from figure \ref{conn} (left columns) with that determined by truncating \eqref{FGapprox} at order $z^2$ (right columns,  with ``Pred.'' and ``Approx.'' referring to analytic and numerical results respectively). The numerical precision is shown when it falls below three significant figures, estimated by comparing results for 100 and 150 lattice points and for fitting different ranges of $L$ depending on the dimension.
}
\label{smallLS}
\end{figure}

\subsection{Wide slabs}
\label{confwide}

For $L \gg b$, the entangling surface is given by two disconnected disks each localized near $x=\pm L$. As in section \ref{ATFDopposed}, the entropy depends on $L$ only through the local behavior of $b(x)$ near $x= \pm L$.  Furthermore, the contribution from each surface is just the entropy one would compute for the corresponding half-space $x > L$ or  $x  < - L$.  For simplicity we thus focus below on this notion of `half space entropy' and choose ${\cal R}_{CFT}$ to be the region $x>\pm L$. Note that our geometry ends at $z=\tilde b$, with the extremal surface obeying the boundary condition of regularity \reef{bcCB}.

The detailed computations are much as in section \ref{ATFDopposed}, so we simply display the results. The area of the disconnected surface can be written in terms of the dimensionless functions described in \reef{genC} and \reef{genC2} after replacing ${\sf A_c}(L/b_0)$ with ${\sf A_d}$.  We compute the zeroth order coefficients analytically, but the second order coefficients require numerics.  Half-space entropy results for $3 \leq d \leq 7$ are tabulated in figure \ref{crossAS} using our by-now standard notation.
 \begin{figure}[h!]
\centering
\begin{tabular}{c || c||c | c }
$d$ &  ${\sf A_d}^{(0)}$& $ {\sf A_d}^{({ b'}^2)}$ & $ {\sf A_d}^{( b\,  b'')}$\\
\hline
$3 $ & $ -0.667 $ & $ -0.0882$& $ 0.0882 $\\
$4 $ & $ -0.354 $ & $-0.0424$& $0.0283 $\\
$5 $ & $ -0.232 $ & $ -0.0875 $& $ 0.0437 $\\
$6 $ & $ -0.167 $ & $ -0.135$& $ 0.0540$\\
$7 $ & $ -0.126 $ & $-0.158 $& $ 0.0527 $
\end{tabular}

\caption{The coefficients ${\sf A_d}^{(0)}$,  ${\sf A_d}^{({b'}^2)}$, and ${\sf A_d}^{(b\, b'')}$. The numerical precision is around six significant figures, estimated by comparing results for $100$ and $150$ lattice points.
}
\label{crossAS}
\end{figure}

\subsection{Phase transition}
\label{GSCphase}

Finally, we turn to the effect of adiabatic variations on the critical value $L_{crit}$ at which the dominant phase becomes disconnected.  As in section \ref{ATFDphase}, we do so by comparing the area of the connected surface (figure \ref{conn}) with the area of the disconnected surface evaluated at $x =\pm L$ (figure \ref{crossAS}). Since the phase transition occurs at $L \ll b/(\epsilon b')$, we again expand $b(x)$ in a Taylor's series to compute ${\sf A}_d$.   The second-order coefficients of of $L_{crit}$ are again given by \eqref{L2} with the replacements $2 A_{ren} \rightarrow {\sf A}_c$, $\hat A_{ren} \rightarrow {\sf A}_d$. We determine $\lc$ numerically to second order, and display these results in figure \ref{soliton} using the notation of \reef{lcritcoeff}.

\begin{figure}[h!]
\centering
\begin{tabular}{c||c||c |c }
$d$ & $L^{(0)}_\text{crit}/ b_0$& $   {\sf L}^{({b'_0}^2)} $& $ {\sf L}^{(b_0b''_0)}$\\
\hline
$3 $ & $ 0.249 $ & $ -0.0475\pm 0.0002 $ &  $0.0116\pm 0.0002$ \\
$4 $ & $ 0.217 $ & $-0.0694 $ & $ 0.287$ \\
$5 $ & $ 0.191 $ & $ -0.107 \pm 0.004$ & $ 0.0233$ \\
$6 $ & $ 0.170 $ & $ -0.167 \pm 0.017 $ & $ 0.0194$ \\
$7 $ & $ 0.152 $ & $-0.237 \pm 0.036 $ & $0.0157 $
\end{tabular}

\caption{
Numerical values of $L_\text{crit}/b_0$ and the coefficients ${\sf L}^{({b'_0}^2)} $ and  $ {\sf L}^{(b_0b''_0)}$ from \eqref{L2} for the RT phase transition for slabs $|x|\le L$ in our confined ground state with $3\le d \le 7$. The numerical precision is shown when it falls below three figures, estimated by comparing results for $100$ and $150$ lattice points.
}
\label{soliton}
\end{figure}

%%%%%%%%%%%%%%%%%%%%%%%%%%%%%%%%%%%%%%%%%%%%%%%%%%%%%%%%%%%%%%%%%%%%%%%%%%%%%%%%%%%%%%%%%%%%%%%%%
%%%%%%%%%%%%%%%%%%%%%%%%%%%%%%%%%%%%%%%%%%%%%%%%%%%%%%%%%%%%%%%%%%%%%%%%%%%%%%%%%%%%%%%%%%%%%%%%%

\section{Discussion}
\label{discuss}

In the above work, we computed the leading (second order) effects of inhomogeneities on the holographic entropy of slab-shaped regions defined by $|x-x_0| \le L$. We studied thermofield-double states on spacetimes where the redshift changes slowly with position, and the ground states of certain confining theories with corresponding slow changes in the confinement scale.  In each case, we studied the effect on the length scale $L_{crit}$ associated with a Ryu-Takayanagi phase transition.  Most of our results were numerical, though the special case $d=2$ (AdS$_{3}$) was treated analytically in appendix \ref{2+1}.  In higher dimensions, some analytic results were also available in special limits and were used to check our numerics.

For the thermofield double, $L_{crit}$ is a measure of the non-locality of entanglements between opposite CFTs.   The second-order coefficients (figure \ref{lcrit}) governing the response of $L_{crit}$ to inhomogeneities turn out to be numerical small.   Some insight  as to why is provided by the analytic $d=2$ treatment of appendix \ref{2+1}, which shows these coefficients to be proportional to $(L_{crit}/b)^3$.  So the small values of $L_{crit}/b$ lead to even smaller coefficients ${\sf L}^{(b'_0{}^2)}$,  ${\sf L}^{(b_0b''_0)}$.

The coefficients shown in figure \ref{lcrit} display highly non-trivial structure with respect to the dimension $d$.  For $d \le 3$, gradients decrease $L_{crit}$, while they increase $L_{crit}$ for $d \ge 4$.  This remains true whether one studies the local response to $b_0'$ or the average change over all $x$.  The former is precisely the sign of ${\sf L}^{(b'_0{}^2)}$ in figure \ref{lcrit}.   But averaging $L^{(2)}_{crit}$ over $x$ allows one to use either periodic boundary conditions or $b \rightarrow constant$ as $x \rightarrow \pm \infty$ to integrate $b^2 b''$ by parts, giving a positive-definite quantity multiplied by $({\sf L}^{(b'_0{}^2)} - 2{\sf L}^{(b_0b''_0)})$.  It turns out that both change sign between $d=3$ and $d=4$.  Interestingly, it is the large $d$ behavior that corresponds to the naive expectation that that the response is given by averaging $b(x)$ over a scale $|x -x_0| \lesssim b$, as such averaging would decrease $L_{crit}$ near a maximum of $b(x)$ and thus require ${\sf L}^{(b_0b''_0)} < 0$.  This is the opposite sign to that found analytically for $d=2$ in appendix \ref{2+1}.

One also notes that the coefficients ${\sf L}^{(b_0b''_0)}$ are not monotonic with $d$, but appear to have a local minimum near $d=6$.    In contrast, ${\sf L}^{(b'_0{}^2)}$ appears to be monotonic in $d$ but is also highly non-uniform.  In particular, while most cases exhibit a clear increase in value with $d$, the coefficients for $d=5$ and $d=6$ are remarkably close.  The in-depth analysis of numerical convergence in appendix \ref{error} appear to confirm that these features are real and are not just numerical artifacts. It would be useful to have an analytic understanding of these dimension-dependent features; the large $d$ limit may be worth particular study.

In contrast, the response of our confining ground states  is both larger and more uniform in $d$;  figure \ref{soliton} shows no changes of signs.   It is nevertheless interesting that gradients  -- either local or averaged -- always decrease $L_{crit}$.  This is naturally understood as a corresponding decrease in the length scale characterizing confinement.  But comparing our results with \cite{metricpaper} challenges this interpretation.  For $d \le 5$, \cite{metricpaper} found that the gradients {\it decrease} the tension of flux tubes aligned in their direction, while the increase of tension one would expect from a decrease in the confinement length scale occurred only for $d \ge 6$.  Furthermore, for $d>3$ it found that gradients always raised the negative energy of the confining ground state -- a result naturally associated with a larger confinement length scale.  The main conclusion appears to be that confinement is not generally characterized by a single scale, but that changes in different confinement-related phenomenon under small perturbations are often uncorrelated.  It would be interesting to develop more analytic understanding of such effects, and also to determine to what extent our results apply to other systems with spatially-varying confinement scale such as those that might be constructed in a condensed matter laboratory.

\section*{Acknowledgements}
It is a pleasure to thank Eric Mefford, Sebastian Fischetti, William Kelly, and Jorge Santos for useful discussions.  This work was supported in part by the Simons Foundation and by funds from the University of California.

%%%%%%%%%%%%%%%%%%%%%%%%%%%%%%%%%%%%%%%%%%%%%%%%%%%%%%%%%%%%%%%%%%%%%%%%%%%%%%%%%%%%%%%%%%%%%%%%%
%%%%%%%%%%%%%%%%%%%%%%%%%%%%%%%%%%%%%%%%%%%%%%%%%%%%%%%%%%%%%%%%%%%%%%%%%%%%%%%%%%%%%%%%%%%%%%%%%

\appendix

\section{Adiabatic Thermofield Doubles in 1+1 Dimensions}
\label{2+1}

Holographic 1+1 CFTs have asymptotically AdS${}_3$ bulk duals.
Due to the lack of local gravitational degrees of freedom in 2+1 dimensions, all complete asymptotically locally AdS spacetimes are diffeomorphic to global AdS$_3$ (or to a quotient thereof).  This fact greatly simplifies the associated minimal surfaces, allowing us to compute properties of adiabatic thermofield-double analytically for $d=2$.  We do so here in an attempt to gain insight into our numerical results, and also as a check on our numerics.

For $d=2$, the zeroth order ansatz \reef{AdS-Soliton} becomes simply
\ban{
ds^2=\frac1 {z^2} \left[dz^2+b^2 {\left(1-\frac{z^2}{{b}^2}\right)^2 d\theta^2 + \left(1+\frac {z^2}{b^2}\right)^{2}} dx^2\right]\ \ . \label{BTZ}
}
As shown in appendix A of \cite{metricpaper}, the second order corrections are
\ban{
\gp{\theta\theta}&= \frac{z^2 \left(b^2-z^2\right) {b'}^2}{2\, b^2}\mathbreak
\gp{xx}&=\frac{z^2 \left(b^2+z^2\right) \left(2 \, b \,b''-{b'}^2\right)}{2 b^4} \ \ .
}
Using \eqref{floor}, this places the horizon at
\ban{
\label{2dhorizon}
z_H= b +\epsilon^2 \, \frac 18  \, b \, {b'}^2 + O(\epsilon^4).
}

We can now compute various entropies.  Taking ${\cal R}_{CFT}$ to be the half space $x > 0$ in the union of the two CFTs, the equation of motion for the first order correction $x^{(1)}(z)$ to the entangling surface reduces to
\ban{
0&= \left(b^2-3 z^2\right) \, \partial_z x^{(1)}(z)-z \left(b^2+z^2\right) \partial_z^2 x^{(1)}(z),
}
and the boundary conditions become
\ban{
x^{(1)}(0)&=0\mathbreak
x^{(1)}(b)&=-\frac 14 b\, b' \ \ .
}
The solution is given by
\ban{\label{x1sold2}
x^{(1)}(z)& = -\frac{b\, b'\,  z^2}{2 \left(b^2+z^2\right)}\ \ .
}
Comparing \eqref{x1sold2} to our numerics for $d=2$ gives agreement to one part in $10^{16}$. Turning now to the renormalized entropy, using \eqref{2dhorizon} we find that the second order contribution coming from integrating the zeroth order surface over the region $z \in[b, z_H]$ precisely cancels the second order contribution associated with the first-order shift of extremal surface within the zeroth order background. As these were the only possible contributions to this order, in agreement with our numerics we find that the full second order contribution vanishes exactly.

We may also analytically compute the entropy of a strip (analogous to our slabs in higher dimensions). We take the strip to be thin compared to the adiabatic scale ($L\ll b/\epsilon b'$). Solving the equations of motion gives
\ban{\label{2dentsurf}
z^{(0)}(x)&= b_0 \sqrt{\frac{\cosh \frac{2 L }{b_0}-\cosh \frac{2 x}{b_0}}{\cosh \frac{2 x}{b_0}+\cosh \frac{2 L }{b_0}}}\mathbreak
z^{(1)}(x)&=b_0' \frac{ \left(-2 \left(b_0^2-2 x^2+2 L ^2\right) \sinh \frac{2 x}{b_0} \cosh \frac{2 L }{b_0}+2 b_0 x \cosh \frac{4 L }{b_0}+b_0 \left(b_0 \sinh \frac{4 x}{b_0}-2 x \cosh \frac{4 x}{b_0}\right)\right)}{4 \sqrt{\cosh \frac{2 L }{b_0}-\cosh \frac{2 x}{b_0}} \left(\cosh \frac{2 x}{b_0}+\cosh \frac{2 L }{b_0}\right){}^{3/2}}.
}
The numerically derived surfaces agree with the above to one part in $10^{14}$ to zeroth order and one part in $10^{7}$ to first order. Computing the entanglement entropy gives
\ban{
A^{(0)}_\text{ren} &= 2 \log \sinh \frac{2 L}{b_0} \mathbreak
A^{(2)}_\text{ren} &= \left( -\frac{L^2}{b_0^2} +\frac 43\frac{ L^3}{ b_0^3} \coth \frac{2L}{b_0}\right){b_0'}^2 + \left( \frac{L^2}{b_0^2} -\frac 23\frac{ L^3}{ b_0^3} \coth \frac{2L}{b_0}\right){b_0''} .
\label{2danalytic}
}
Comparing this result to our $d=2$ numerics shows discrepancies only at the level of one part in $10^{4}$ level  for the coefficient of ${b_0'}^2$ and one part in $10^{15}$ for the coefficient of $b_0''$.

With these expressions for the area, we can compute the location of the phase transition between vanishing and non-vanishing mutual information to second order. To zeroth order, for the half space entangling surface we have $\hat {\sf A}^{(0)}=0$ so from \eqref{2danalytic} $\hat A_{\text{ren}}^{(0)}  = A_{\text{ren}}^{(0)} $ gives
\ban{
\label{2dPT}
L_\text{crit} ^{(0)}= \frac {b_0}{2} \sinh^{-1}(1) \, .
}
At first order it is manifest  that $A_{\text{ren}}^{(1)}=0$.  In contrast,  keeping in mind the renormalization prescription \reef{d24}, the area of the entangling surface for half space $x<L$ does have a first order correction.  But it is canceled by the corresponding correction to the entangling surface for $x>-L$, so the $O(\epsilon)$ correction $L_\text{crit}^{(1)}$ to $L_{crit}$ vanishes.

However, at second order we find
\ban{
\hat A_\text{ren}^{(2)} = -\frac \ell 2 \frac{L^2}{b_0^2} {b_0'}^2 + \frac \ell 2 \frac{L^2}{b_0} {b_0''} \, .
}
Comparing with \reef{2danalytic} and using \eqref{L2} yields
\ban{
\label{L22d}
L_\text{crit}^{(2)} &= -\frac{b_0}{48} \sinh ^{-1}(1)^3 (2 {b_0'}^2- b_0 b_0'')\,.
}
This result agrees with the results in figure \ref{lcrit} to one part in $10^4$.

As a final check on our $d=2$ results we can solve for the diffeomorphism taking
$g^{(0)}_{\mu\nu}$ with constant $b_0$ to $\tilde g^{(0)}_{\mu\nu} : = g^{(0)}_{\mu\nu}+\epsilon^2\,g^{(2)}_{\mu\nu}$.
Working near $x =0$, we find that the correct diffeomorphism beomes
\ban{
 \tilde z&= z+ \epsilon \, z \,  \frac{x\, b_0'}{b_0}+\epsilon ^2\, z\, \frac{  \left({2 x^2 \left(b_0^2+z^2\right) \left({b_0'}{}^2+b_0\,  b_0''\right)}-z^2 \, {b_0^2}\, {b_0'}{}^2\right)}{4 {b_0^2}\, \left(b_0^2+z^2\right)}+O(\epsilon^3)\mathbreak
\tilde x&=x+\epsilon\, \frac{  b_0' \left(b_0^2 (x^2-z^2)+x^2 z^2\right)}{2 b_0 \left(b_0^2+z^2\right)}+ \epsilon ^2\,\frac{x  \left(b_0^2 \left(x^2-3 z^2\right)+x^2 z^2\right) \left({b_0'}{}^2+b_0 b_0''\right)}{6 b_0^2 \left(b_0^2+z^2\right)}+O(\epsilon^3)
\label{2ddiff},}
which indeed takes the entangling surfaces of global AdS$_3$ to \eqref{2dentsurf} as desired.  One may also check that \eqref{2ddiff} maps the phase transition for $b(x)=constant$ (given by \eqref{2dPT}) to the value specified by \eqref{L22d}.

%%%%%%%%%%%%%%%%%%%%%%%%%%%%%%%%%%%%%%%%%%%%%%%%%%%%%%%%%%%%%%%%%%%%%%%%%%%%%%%%%%%%%%%%%%%%%%%%%
%%%%%%%%%%%%%%%%%%%%%%%%%%%%%%%%%%%%%%%%%%%%%%%%%%%%%%%%%%%%%%%%%%%%%%%%%%%%%%%%%%%%%%%%%%%%%%%%%

\section{Estimation of Numerical Uncertainty}
\label{error}

We have used two distinct methods to estimate the numerical uncertainty of our results. First, for the majority of the tables we merely make a rough estimate by computing a particular coefficient using a variety numerical parameters.  We then take the approximate error to be given by the standard deviation of this set. For example for the $L\gg b_0$ scaling of figure \ref{largeL}, we compare values calculated using $100$ and $150$ lattice points and for fitting intervals $L/b_0 \in [40,50]$ and $L/b_0 \in [50,60]$. The estimated error is the standard deviation of this four point data set.  The value displayed in the table is the mean.

However, as noted in the main text, the values tabulated in figure \ref{lcrit} are rather less uniform than one might expect.  As a result, we now take extra care to analyze the numerical results reported there.  After investigating the possible sources of error by varying the precision of different parts of the computation, we find the dominant error (by far) to come from using a finite number $N$ of lattice points in the interval $[-L,L]$.  We now study how our results change with $N$.

\begin{figure}[h!]
\centering
\subfloat[][$d=5, \, p=7/4$]{\includegraphics[width=0.45\textwidth]{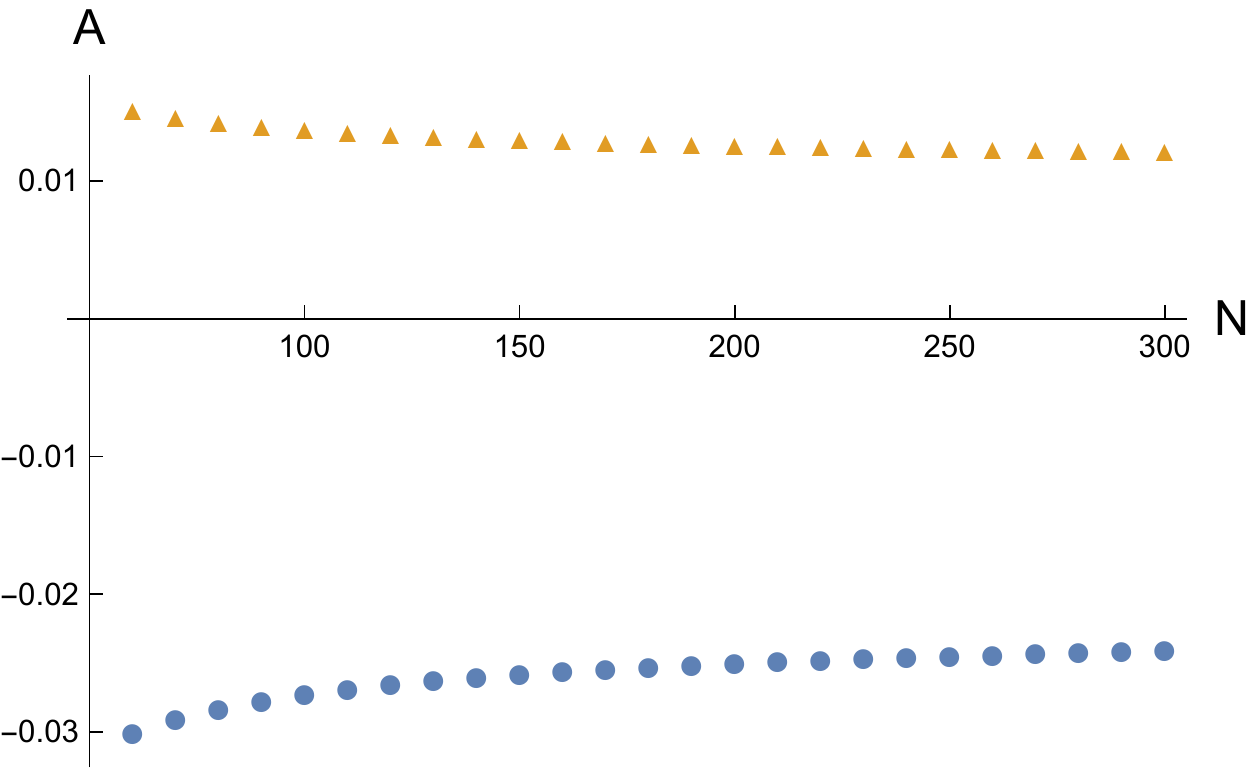}}\qquad
\subfloat[][$d=6,\, p=7/4$]{\includegraphics[width=0.45\textwidth]{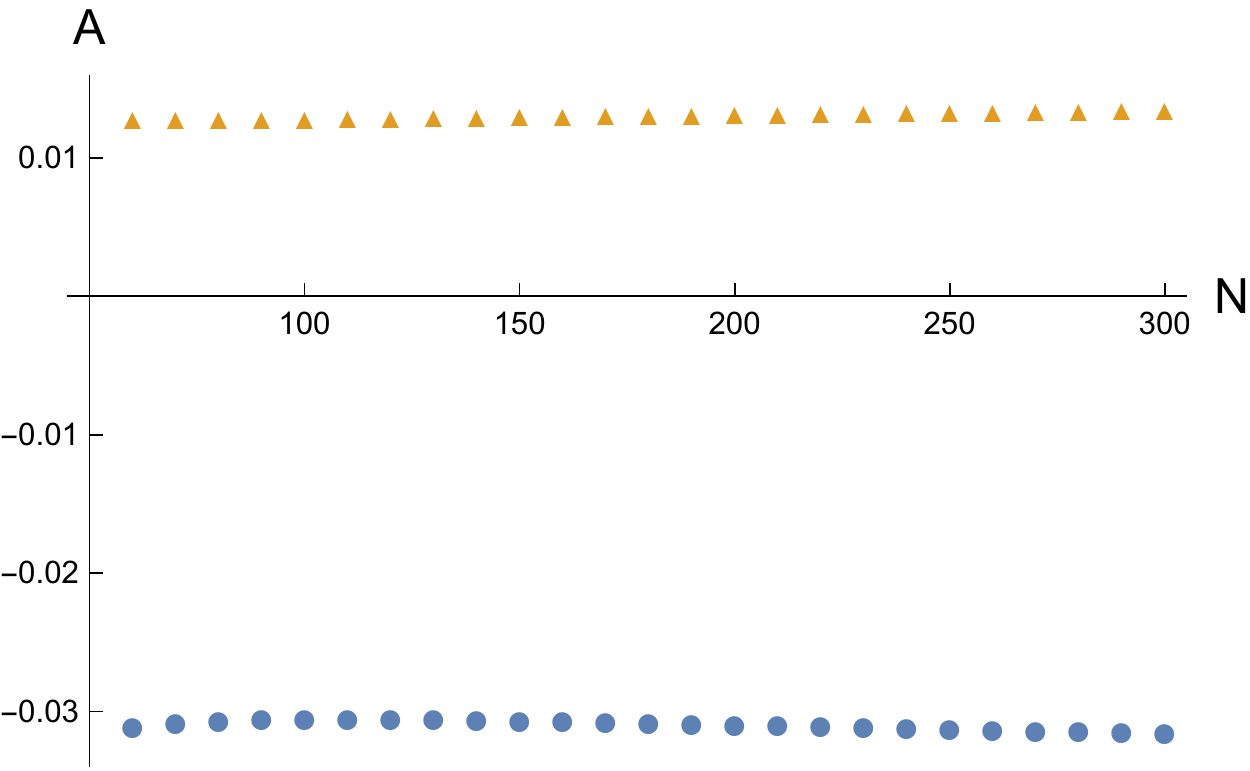}}\\
\subfloat[][$d=7, \,p=13/8$]{\includegraphics[width=0.65\textwidth]{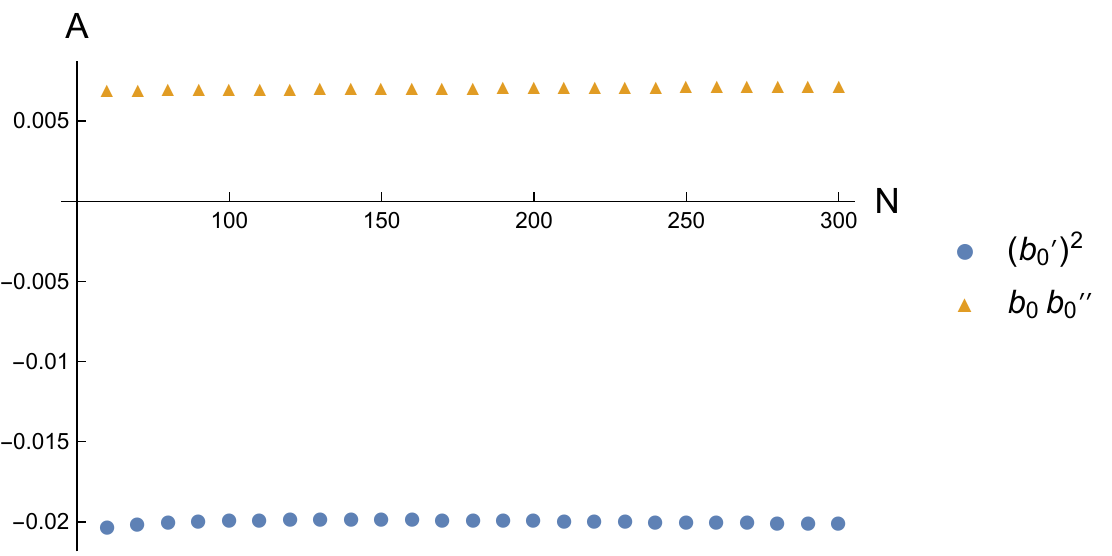}}
\caption{ Plots of $D_N$ as defined in \reef{errorscaling} vs. N with $d=5,6,7$ for the ${b_0'}^2$ and $b_0 b_0''$ coefficients (triangles and disks respectively). We choose $p$ so that the datasets are either flat or slowly approaching zero.
}
\label{errorplots}
\end{figure}

We first compute $L_\text{crit}$ using $N=[50,300]$ lattice points in steps of $10$. Next, we approximate the function $\deriv{L_\text{crit}}N$ by choosing an appropriate $p$ so that the data
\ban{
D_N= \frac 1{10}\left[L_\text{crit}(N) -L_\text{crit}(N+10)\right]N^p
\label{errorscaling}
}
appears constant to the eye. See figure \ref{errorplots} for examples.  We then compute the average $\bar D$ of $D_N$ over the data set and model our results by
 \ban{
\deriv{L_\text{crit}(N)}N = \bar D\, N^p\, .
\label{Nmodel} }
Given \eqref{Nmodel}, we can compute $\Delta(N_0) = L_\text{crit}(N_0)-L_\text{crit}(\infty)$.  We have reported the values $\Delta(N_0)$ for $N_0 =300$ as the numerical uncertainties in figure \ref{errorplots}.  Though we do not fully understand the particular values of $p$ found in this way, we believe this to be a conservative estimate of our errors (especially when $D_N$ clearly decreases). We display $\bar D$ as well as the determined value of $p$ for $2\leq d \leq 8$ in figure \ref{pvalues}.

\begin{figure}[h!]
\centering
\begin{tabular}{c||c|c|c}
$d$& $\bar D^{(b_0'{}^2)}$ & $\bar D^{(b_0b_0'')}$ & $p$\\
\hline
2& $0.00301 $&$1.94 \times 10^{-8}$ &2.75\\
3& $-0.0819 $&$0.0365$ &2\\
4& $-0.0224 $&$0.0116$ &1.75\\
5& $-0.0256 $&$0.0130$ &1.75\\
6& $-0.0309 $&$0.0131$ &1.75\\
7& $-0.0199 $&$0.00710$ &1.625\\
8& $-0.0128 $&$0.00391$ &1.5
\end{tabular}
\caption{We display the estimated values of $\bar D$ for each of the coefficients $b_0'{}^2$ and $b_0b_0''$ and $p$ for $2\leq d \leq 8$.
}
\label{pvalues}
\end{figure}

%%%%%%%%%%%%%%%%%%%%%%%%%%%%%%%%%%%%%%%%%%%%%%%%%%%%%%%%%%%%%%%%%%%%%%%%%%%%%%%%%%%%%%%%%%%%%%%%%
%%%%%%%%%%%%%%%%%%%%%%%%%%%%%%%%%%%%%%%%%%%%%%%%%%%%%%%%%%%%%%%%%%%%%%%%%%%%%%%%%%%%%%%%%%%%%%%%%

\bibliographystyle{jhep}
	\cleardoublepage
\phantomsection
\renewcommand*{\bibname}{References}

\bibliography{references}

\providecommand{\href}[2]{#2}\begingroup\raggedright\begin{thebibliography}{10}

\bibitem{maldTFD}
J.~M. Maldacena, {\it {Eternal black holes in anti-de Sitter}},  {\em JHEP}
  {\bf 04} (2003) 021, [\href{http://arxiv.org/abs/hep-th/0106112}{{\tt
  arXiv:hep-th/0106112}}].

\bibitem{VanRaamsdonk:2010pw}
M.~Van~Raamsdonk, {\it {Building up spacetime with quantum entanglement}},
  {\em Gen. Rel. Grav.} {\bf 42} (2010) 2323--2329,
  [\href{http://arxiv.org/abs/1005.3035}{{\tt arXiv:1005.3035}}]. [Int. J. Mod.
  Phys.D19,2429(2010)].

\bibitem{Czech:2012be}
B.~Czech, J.~L. Karczmarek, F.~Nogueira, and M.~Van~Raamsdonk, {\it {Rindler
  Quantum Gravity}},  {\em Class. Quant. Grav.} {\bf 29} (2012) 235025,
  [\href{http://arxiv.org/abs/1206.1323}{{\tt arXiv:1206.1323}}].

\bibitem{Maldacena:2013xja}
J.~Maldacena and L.~Susskind, {\it {Cool horizons for entangled black holes}},
  {\em Fortsch. Phys.} {\bf 61} (2013) 781--811,
  [\href{http://arxiv.org/abs/1306.0533}{{\tt arXiv:1306.0533}}].

\bibitem{Marolf:2015vma}
D.~Marolf, H.~Maxfield, A.~Peach, and S.~F. Ross, {\it {Hot multiboundary
  wormholes from bipartite entanglement}},  {\em Class. Quant. Grav.} {\bf 32}
  (2015), no.~21 215006, [\href{http://arxiv.org/abs/1506.04128}{{\tt
  arXiv:1506.04128}}].

\bibitem{witten}
E.~Witten, {\it {Anti-de Sitter space, thermal phase transition, and
  confinement in gauge theories}},  {\em Adv. Theor. Math. Phys.} {\bf 2}
  (1998) 505--532, [\href{http://arxiv.org/abs/hep-th/9803131}{{\tt
  arXiv:hep-th/9803131}}].

\bibitem{metricpaper}
D.~Marolf and J.~Wien, {\it {Holographic confinement in inhomogeneous
  backgrounds}},  {\em JHEP} {\bf 08} (2016) 015,
  [\href{http://arxiv.org/abs/1605.02804}{{\tt arXiv:1605.02804}}].

\bibitem{FG}
S.~Bhattacharyya, V.~E. Hubeny, S.~Minwalla, and M.~Rangamani, {\it {Nonlinear
  Fluid Dynamics from Gravity}},  {\em JHEP} {\bf 02} (2008) 045,
  [\href{http://arxiv.org/abs/0712.2456}{{\tt arXiv:0712.2456}}].

\bibitem{FGref1}
V.~E. Hubeny, S.~Minwalla, and M.~Rangamani, {\it {The fluid/gravity
  correspondence}},  in {\em {Black holes in higher dimensions}}, pp.~348--383,
  2012.
\newblock \href{http://arxiv.org/abs/1107.5780}{{\tt arXiv:1107.5780}}.

\bibitem{FGref2}
M.~Rangamani, {\it {Gravity and Hydrodynamics: Lectures on the fluid-gravity
  correspondence}},  {\em Class. Quant. Grav.} {\bf 26} (2009) 224003,
  [\href{http://arxiv.org/abs/0905.4352}{{\tt arXiv:0905.4352}}].

\bibitem{rt1}
S.~Ryu and T.~Takayanagi, {\it {Holographic derivation of entanglement entropy
  from AdS/CFT}},  {\em Phys. Rev. Lett.} {\bf 96} (2006) 181602,
  [\href{http://arxiv.org/abs/hep-th/0603001}{{\tt arXiv:hep-th/0603001}}].

\bibitem{rt2}
S.~Ryu and T.~Takayanagi, {\it {Aspects of Holographic Entanglement Entropy}},
  {\em JHEP} {\bf 08} (2006) 045,
  [\href{http://arxiv.org/abs/hep-th/0605073}{{\tt arXiv:hep-th/0605073}}].

\bibitem{head}
M.~Headrick, {\it {Entanglement Renyi entropies in holographic theories}},
  {\em Phys. Rev.} {\bf D82} (2010) 126010,
  [\href{http://arxiv.org/abs/1006.0047}{{\tt arXiv:1006.0047}}].

\bibitem{fur06a}
D.~V. Fursaev, {\it {Proof of the holographic formula for entanglement
  entropy}},  {\em JHEP} {\bf 09} (2006) 018,
  [\href{http://arxiv.org/abs/hep-th/0606184}{{\tt arXiv:hep-th/0606184}}].

\bibitem{Lewkowycz:2013nqa}
A.~Lewkowycz and J.~Maldacena, {\it {Generalized gravitational entropy}},  {\em
  JHEP} {\bf 08} (2013) 090, [\href{http://arxiv.org/abs/1304.4926}{{\tt
  arXiv:1304.4926}}].

\bibitem{Haehl:2014zoa}
F.~M. Haehl, T.~Hartman, D.~Marolf, H.~Maxfield, and M.~Rangamani, {\it
  {Topological aspects of generalized gravitational entropy}},  {\em JHEP} {\bf
  05} (2015) 023, [\href{http://arxiv.org/abs/1412.7561}{{\tt
  arXiv:1412.7561}}].

\bibitem{Hubeny:2007xt}
V.~E. Hubeny, M.~Rangamani, and T.~Takayanagi, {\it {A Covariant holographic
  entanglement entropy proposal}},  {\em JHEP} {\bf 07} (2007) 062,
  [\href{http://arxiv.org/abs/0705.0016}{{\tt arXiv:0705.0016}}].

\bibitem{Wall:2012uf}
A.~C. Wall, {\it {Maximin Surfaces, and the Strong Subadditivity of the
  Covariant Holographic Entanglement Entropy}},  {\em Class. Quant. Grav.} {\bf
  31} (2014), no.~22 225007, [\href{http://arxiv.org/abs/1211.3494}{{\tt
  arXiv:1211.3494}}].

\bibitem{Graham:1999pm}
C.~R. Graham and E.~Witten, {\it {Conformal anomaly of submanifold observables
  in AdS / CFT correspondence}},  {\em Nucl. Phys.} {\bf B546} (1999) 52--64,
  [\href{http://arxiv.org/abs/hep-th/9901021}{{\tt arXiv:hep-th/9901021}}].

\bibitem{StateDep}
D.~Marolf and A.~C. Wall, {\it {State-Dependent Divergences in the Entanglement
  Entropy}},  \href{http://arxiv.org/abs/1607.01246}{{\tt arXiv:1607.01246}}.

\bibitem{Jacobson:1993vj}
T.~Jacobson, G.~Kang, and R.~C. Myers, {\it {On black hole entropy}},  {\em
  Phys. Rev.} {\bf D49} (1994) 6587--6598,
  [\href{http://arxiv.org/abs/gr-qc/9312023}{{\tt arXiv:gr-qc/9312023}}].

\bibitem{Dong:2013qoa}
X.~Dong, {\it {Holographic Entanglement Entropy for General Higher Derivative
  Gravity}},  {\em JHEP} {\bf 01} (2014) 044,
  [\href{http://arxiv.org/abs/1310.5713}{{\tt arXiv:1310.5713}}].

\bibitem{Miao:2014nxa}
R.-X. Miao and W.-z. Guo, {\it {Holographic Entanglement Entropy for the Most
  General Higher Derivative Gravity}},  {\em JHEP} {\bf 08} (2015) 031,
  [\href{http://arxiv.org/abs/1411.5579}{{\tt arXiv:1411.5579}}].

\bibitem{Taylor:2016aoi}
M.~Taylor and W.~Woodhead, {\it {Renormalized entanglement entropy}},
  \href{http://arxiv.org/abs/1604.06808}{{\tt arXiv:1604.06808}}.

\bibitem{Morrison:2012iz}
I.~A. Morrison and M.~M. Roberts, {\it {Mutual information between thermo-field
  doubles and disconnected holographic boundaries}},  {\em JHEP} {\bf 07}
  (2013) 081, [\href{http://arxiv.org/abs/1211.2887}{{\tt arXiv:1211.2887}}].

\bibitem{Headrick:2010zt}
M.~Headrick, {\it {Entanglement Renyi entropies in holographic theories}},
  {\em Phys. Rev.} {\bf D82} (2010) 126010,
  [\href{http://arxiv.org/abs/1006.0047}{{\tt arXiv:1006.0047}}].

\bibitem{Hartman:2013qma}
T.~Hartman and J.~Maldacena, {\it {Time Evolution of Entanglement Entropy from
  Black Hole Interiors}},  {\em JHEP} {\bf 05} (2013) 014,
  [\href{http://arxiv.org/abs/1303.1080}{{\tt arXiv:1303.1080}}].

\bibitem{dlmf}
``{NIST Digital Library of Mathematical Functions}.'' http://dlmf.nist.gov/,
  Release 1.0.10 of 2015-08-07.

\bibitem{numerics}
O.~J.~C. Dias, J.~E. Santos, and B.~Way, {\it {Numerical Methods for Finding
  Stationary Gravitational Solutions}},
  \href{http://arxiv.org/abs/1510.02804}{{\tt arXiv:1510.02804}}.

\bibitem{Klebanov:2007ws}
I.~R. Klebanov, D.~Kutasov, and A.~Murugan, {\it {Entanglement as a probe of
  confinement}},  {\em Nucl. Phys.} {\bf B796} (2008) 274--293,
  [\href{http://arxiv.org/abs/0709.2140}{{\tt arXiv:0709.2140}}].

\bibitem{Nishioka:2006gr}
T.~Nishioka and T.~Takayanagi, {\it {AdS Bubbles, Entropy and Closed String
  Tachyons}},  {\em JHEP} {\bf 01} (2007) 090,
  [\href{http://arxiv.org/abs/hep-th/0611035}{{\tt arXiv:hep-th/0611035}}].

\end{thebibliography}\endgroup

\end{document}